\documentclass[twocolumn,showpacs,preprintnumbers,amsmath,amssymb,eqsecnum]{revtex4}
\usepackage{dcolumn}
\usepackage{bm}
\usepackage[dvips]{graphicx}
\usepackage{wrapfig}
\usepackage{psfrag}
\begin{document}

\def\sh{\mathop{\rm sh}\nolimits}
\def\ch{\mathop{\rm ch}\nolimits}
\def\var{\mathop{\rm var}}\def\exp{\mathop{\rm exp}\nolimits}
\def\Re{\mathop{\rm Re}\nolimits}
\def\Sp{\mathop{\rm Sp}\nolimits}
\def\kp{\mathop{\text{\ae}}\nolimits}
\def\bk{{\bf {k}}}
\def\bp{{\bf {p}}}
\def\bq{{\bf {q}}}
\def\lra{\mathop{\longrightarrow}}
\def\Const{\mathop{\rm Const}\nolimits}
\def\sh{\mathop{\rm sh}\nolimits}
\def\ch{\mathop{\rm ch}\nolimits}
\def\var{\mathop{\rm var}}

\def\Re{\mbox {Re}}
\newcommand{\Z}{\mathbb{Z}}
\newcommand{\R}{\mathbb{R}}
\def\mK{\mathop{{\mathfrak {K}}}\nolimits}
\def\mR{\mathop{{\mathfrak {R}}}\nolimits}
\def\mv{\mathop{{\mathfrak {v}}}\nolimits}
\def\mV{\mathop{{\mathfrak {V}}}\nolimits}
\def\mD{\mathop{{\mathfrak {D}}}\nolimits}
\def\mN{\mathop{{\mathfrak {N}}}\nolimits}
\newcommand{\ccm}{{\cal M}}
\newcommand{\cE}{{\cal E}}
\newcommand{\cV}{{\cal V}}
\newcommand{\cI}{{\cal I}}
\newcommand{\cR}{{\cal R}}
\newcommand{\cK}{{\cal K}}
\newcommand{\cH}{{\cal H}}

\def\br{\mathop{{\bf {r}}}\nolimits}
\def\bS{\mathop{{\bf {S}}}\nolimits}
\def\bA{\mathop{{\bf {A}}}\nolimits}
\def\bJ{\mathop{{\bf {J}}}\nolimits}
\def\bn{\mathop{{\bf {n}}}\nolimits}
\def\bg{\mathop{{\bf {g}}}\nolimits}
\def\bv{\mathop{{\bf {v}}}\nolimits}
\def\be{\mathop{{\bf {e}}}\nolimits}
\def\bp{\mathop{{\bf {p}}}\nolimits}
\def\bz{\mathop{{\bf {z}}}\nolimits}
\def\bbf{\mathop{{\bf {f}}}\nolimits}
\def\bb{\mathop{{\bf {b}}}\nolimits}
\def\ba{\mathop{{\bf {a}}}\nolimits}
\def\bx{\mathop{{\bf {x}}}\nolimits}
\def\by{\mathop{{\bf {y}}}\nolimits}
\def\br{\mathop{{\bf {r}}}\nolimits}
\def\bs{\mathop{{\bf {s}}}\nolimits}
\def\bH{\mathop{{\bf {H}}}\nolimits}
\def\bk{\mathop{{\bf {k}}}\nolimits}
\def\be{\mathop{{\bf {e}}}\nolimits}
\def\bnul{\mathop{{\bf {0}}}\nolimits}
\def\bq{{\bf {q}}}

\newcommand{\oV}{\overline{V}}
\newcommand{\vkp}{\varkappa}
\newcommand{\os}{\overline{s}}
\newcommand{\opsi}{\overline{\psi}}
\newcommand{\ov}{\overline{v}}
\newcommand{\oW}{\overline{W}}
\newcommand{\oPhi}{\overline{\Phi}}

\def\mI{\mathop{{\mathfrak {I}}}\nolimits}
\def\mA{\mathop{{\mathfrak {A}}}\nolimits}

\def\st{\mathop{\rm st}\nolimits}
\def\tr{\mathop{\rm tr}\nolimits}
\def\sign{\mathop{\rm sign}\nolimits}
\def\d{\mathop{\rm d}\nolimits}
\def\const{\mathop{\rm const}\nolimits}
\def\O{\mathop{\rm O}\nolimits}
\def\Spin{\mathop{\rm Spin}\nolimits}
\def\exp{\mathop{\rm exp}\nolimits}

\def\mI{\mathop{{\mathfrak {I}}}\nolimits}
\def\mA{\mathop{{\mathfrak {A}}}\nolimits}

\def\st{\mathop{\rm st}\nolimits}
\def\tr{\mathop{\rm tr}\nolimits}
\def\sign{\mathop{\rm sign}\nolimits}
\def\d{\mathop{\rm d}\nolimits}
\def\const{\mathop{\rm const}\nolimits}
\def\O{\mathop{\rm O}\nolimits}
\def\Spin{\mathop{\rm Spin}\nolimits}
\def\exp{\mathop{\rm exp}\nolimits}

\title{The lattice quantum gravity, its continuum limit and the cosmological constant problem}

\author {S.N. Vergeles}

\affiliation{Landau Institute for Theoretical Physics,
Russian Academy of Sciences,Chernogolovka, Moskow
region, 142432 Russia }

\begin{abstract} Some variant of discrete quantum theory of gravity is constructed
and its naive continual limit is considered. It is shown that this
continual quantum theory of gravity leads to "light" universe
comparatively with the universe in usual quantum theory. Thus in
the theory the cosmological constant problem in inflating Universe
has a natural solution.
\end{abstract}


\maketitle

\section{INTRODUCTION}

Some time ago I have formulated some variant of discrete quantum
gravity \cite{1}, and regularized continual quantum theory of
gravity \cite{2} -- \cite{4}. In the work \cite{5} the arguments
are given in favour of the discrete quantum gravity \cite{1} has
continual limit. The continual limit is described in \cite{2} --
\cite{4}. In the present paper I show that in this theory the
cosmological constant problem has natural solution.

Let's outline shortly the cosmological constant problem
(the reader can find the review of the problem in \cite{6}).

Consider Einstein equation with $\Lambda$-term ($\hbar=c=1$):
\begin{eqnarray}
R_{\mu\nu}-\frac12g_{\mu\nu}R=8\pi
G\,T_{\mu\nu}+\Lambda\,g_{\mu\nu}\,.
\label{introduction10}
\end{eqnarray}
Here
$T_{\mu\nu}$ is energy-momentum tensor of the matter and $\Lambda$
is some constant parameter having the dimension [$cm^{-2}$]. In
the used unit system the Newtonian gravitational constant
\begin{eqnarray}
G\sim l_P^2\sim 2,5\cdot 10^{-66}\,{cm}^2\,,
\label{introduction20}
\end{eqnarray}
and according to experimental data the mean energy density today
is of the order
\begin{gather}
T_{\mu\nu}\sim\rho_1\sim 10^8{cm}^{-4}\longrightarrow
8\pi\,G\,T_{\mu\nu}\sim 5\cdot 10^{-57}{cm}^{-2}\,,
\label{introduction30}
\end{gather}
and
\begin{eqnarray}
\Lambda\sim 10^{-56}{cm}^{-2}\,.
 \label{introduction40}
\end{eqnarray}
Thus, if Einstein equation (\ref{introduction10}) is used for
description of the today dynamics of Universe, the quantities in
its right hand side are of the same order indicated in
(\ref{introduction30}) and (\ref{introduction40}).

Now let us estimate the possible value of the right hand side of
Eq. (\ref{introduction10}) in the framework of canonical quantum
field theory. For simplicity consider energy-momentum tensor in
quantum electrodynamics in flat spacetime:
\begin{eqnarray}
T_{\mu\nu}=-\frac{1}{4\pi}\left(F_{\mu\lambda}F_{\nu}^{\;\;\lambda}-
\frac14\eta_{\mu\nu}F^2\right)+
\nonumber \\
+\frac{i}{2}\left(\opsi\gamma_{(\mu}\nabla_{\nu)}\psi-
\overline{\nabla_{(\nu}\psi}\gamma_{\mu)}\psi\right)\,.
\label{introduction50}
\end{eqnarray}
Casimir effect, predicted in \cite{7} and experimentally verified
in \cite{8}, shows for reality of zero-point energies. Moreover,
the attempts to drop out zero-point energies by appropriate normal
ordering of creating and annihilating operators in energy-momentum
tensor fail for many of reasons (the discussion of this problem
see, for example, in \cite{9}). Thus, at estimating vacuum
expectation value of energy-momentum tensor
(\ref{introduction50}), it should not be performed normal ordering
of creating and annihilating operators in (\ref{introduction50}).
Thus we obtain  for vacuum expectation value of tensor
(\ref{introduction50}) in free theory:
\begin{eqnarray}
\langle T_{\mu\nu}\rangle_0=\int\frac{\d^{(3)}k}{(2\pi)^3}
\left(\frac{k_{\mu}k_{\nu}}{k^0}\bigg|_{k^0=|\bk|}-\right.
\nonumber \\
\left.-\frac{2k_{\mu}k_{\nu}}{k^0}\bigg|_{k^0=\sqrt{m^2+\bk^2}}
\right)\,.
\label{introduction60}
\end{eqnarray}
Here $m$ is the electron mass. The first item in (\ref{introduction60}) gives the
positive contribution but the second item gives the negative
contribution since these items give the boson and fermion
contributions to vacuum energy, respectively. If integration in
(\ref{introduction60}) is restricted by Planck scale, $k_{max}\sim l_P^{-1}$, then
from (\ref{introduction60}) and (\ref{introduction20}) it follows:
\begin{eqnarray}
 8\pi G \langle T_{\mu\nu}\rangle_0\sim l_P^{-2}\sim
 10^{66}\,{cm}^{-2}\,.
 \label{introduction70}
\end{eqnarray}
It is clear that the interaction of fields doesn't changes
qualitatively the estimation (\ref{introduction70}). From (\ref{introduction70}) and (\ref{introduction30})  we see that the
contribution to the righthand side of Eq. (\ref{introduction10})  estimated in the
framework of canonical quantum field theory is larger about
$10^{120}$ times in comparison with the experimental estimations.

It is known that in globally supersimmetric field theories the
vacuum energy is equal to zero \cite{10}. Indeed, in flat spacetime the
anticommutation relations
 \begin{eqnarray}
 \{Q_{\alpha},\,Q^{\dag}_{\beta}\}=(\sigma_{\mu})_{\alpha\beta}{\cal
 P}^{\mu}\,.
  \label{introduction80}
\end{eqnarray}
 take place. Here $Q_{\alpha}$ are supersimmetry generators,
 $\alpha$ è $\beta$ are spinor indexes,
 $\sigma_1,\,\sigma_2$ and $\sigma_3$ are the Pauli matrices, $\sigma_0=1$, and ${\cal
 P}^{\mu}$ is the energy-momentum 4-vector operator.
 If supersimmetry is unbroken, then the vacuum state
 $|0\rangle$ satisfies
 \begin{eqnarray}
Q_{\alpha}|0\rangle=Q_{\alpha}^{\dag}|0\rangle=0\,,
\label{introduction90}
\end{eqnarray}
and (\ref{introduction80}) and (\ref{introduction90}) imply
\begin{eqnarray}
\langle {\cal P}^{\mu}\rangle_0=0\,.
\label{introduction100}
\end{eqnarray}
The equality (\ref{introduction100}) means that the total sum of zero-point energies
in unbroken globally supersimmetric field theories is rigorously
equal to zero.

However, even if supersimmetry takes place on fundamental level,
it is broken on experimentally tested scales. If one assumes that
supersimmetry is unbroken on the scales greater than $k_{SS}\sim
10^{17}\,cm^{-1}\,(\sim 10^3\,GeV)$, then even in this case the
contribution to the right hand side of Eq. (\ref{introduction10}) from zero-point
energies of all normal modes with energies less then $k_{SS}$ will
exceed experimentally known value about $10^{58}$ times.

It follows from the said above that any calculation in the
framework of canonical quantum field theory leads to unacceptable
large vacuum expectation value of energy momentum tensor. The
considered catastrophe isn't solved at present in superstring
theory.

It should be noted here that the problem of cosmological constant
is solved in original theory of G. Volovik \cite{11}. In this theory
the gravitons and other excitations are the quasiparticles in a
more fundamental quantum system --- quantum fluid of the type
${}^3He$ in superfluid phase. Another approach to the problem of
cosmological constant in the frame of M-theory is developed in
works \cite{12}.

This paper is organized as follows. In sect. 2 I define discrete
quantum gravity which has been introduced in \cite{1}. It is shown
qualitatively that this quantum theory display the tendency to
degenerate into macroscopic continual theory and the continual
limit in this discrete theory is possible. Using high-temperature
expansion (which is possible at small times near the singularity)
the important for the following consideration conclusion is made
that the two-point gauge invariant correlators of any fields are
local, i.e. they decrees exponentially in $x$-space. From here and
the dynamics of discrete theory the interesting conclusion about
noncompact packing of field modes in momentum space in the
continual theory is made. In sects. 3 and 4 formulation of the
corresponding continual quantum theory of gravity and its
connection with discrete theory is given \cite{1} -- \cite{5}.
Naturally, the structure of the continual theory is determined by
more fundamental discrete theory. In sect. 5 it is shown that in
the framework of the suggested quantum theory of gravity the
cosmological constant problem can be solved.

\section{Discrete Quantum Gravity }

\subsection{Definition of Action}

Let $\mK$ be a 4-dimensional simplicial complex admitting
gemetrical realization. The definition and required properties of
simplicial complexes can be found in \cite{1}. A detailed theory
of simplicial complexes is given, for example, in
\cite{13} -- \cite{14}. Below instead of "simplicial complex" we say simply
"complex", and the concepts in the following pairs are treated as
synonyms: 0-simplex and vertex; 1-simplex and edge; 2-simplex and
triangle; 3-simplex and tetrahedron. The finite complexes with a
4-disk topology are interesting here. Such complexes have a
boundary $\partial\mK$ which is 3-dimensional complex with
topology of 3-sphere $S^3$. Denote by $\alpha_q,\,\, q=0,1,2,3,4$
the number of q-simplexes of the complex $\mK$. The indexes
$i,j,k,l,\,\ldots$ run through the complex vertices: $a_i,\,\,a_j$
and so on. Two vertices are called adjacent if these two vertices
are the boundary vertices of the same edge.

For convenience I give here the definition of orientation of simplexes and
complexes.

A simplex
\begin{eqnarray}
s^r=\varepsilon(a_0,\,a_1,\,\ldots\,,\,a_r)\equiv
\varepsilon\,a_0\,a_1\,\ldots\,a_r
\label{discr3}
\end{eqnarray}
has an orientation, or is oriented, if every order of its vertices is
assigned a sign "+" or "-", so that orders differing by an odd permutation correspond to opposite signs.
Thus if $\varepsilon=1$ the orientation of simplex (\ref{discr3}) is given by the
orders $(a_0,\,a_1,\,\ldots\,,\,a_r)$ ore $-(a_1,\,a_0,\,\ldots\,,\,a_r)$.
Let $(a_0,\,\ldots\,,\,a_{i-1},\,a_{i+1},\,\ldots\,,\,a_r)$ be the face of a simplex
$s^r$ obtained by eliminating one vertex $a_i$ from the sequence
$a_0,\,a_1,\,\ldots\,,\,a_r$. By definition, the orientation of this face, given by
\begin{eqnarray}
B^{r-1}_i=(-1)^i\,\varepsilon(a_0,\,
\ldots\,,\,a_{i-1},\,a_{i+1},\,\ldots\,,\,a_r)\,,
\label{discr4}
\end{eqnarray}
is called an iduced orientation of the simplex $s^r$.

Denote by $D$ the maximum value of number $r$ in (\ref{discr3}) for all
simplexes of complex. In considered case $D=4$. Thus $D$ is the dimension of complex.
Two oriented $D$-dimensional simplices
$s^D_1$ and $s^D_2$ of a $D$-dimensional simplicial complex are called concordantly
oriented if either the simplices $s^D_1$ and $s^D_2$ have no common $(D-1)$-dimensional faces or the
orientation of their common $(D-1)$-dimensional face $B^{D-1}$ induced by the orientation of the
simplex $s^D_1$ is opposite to the orientation of the same face
$B^{D-1}$ induced by the orientation of the simplex $s^D_2$.
A $D$-dimensional simplicial complex $\mK$ is called orientable
if there exists such an orientation for all its $D$-dimensional simplices
that any pair of its $D$-dimensional simplices is concordantly oriented.
The concordant orientation of $D$-dimensional simplices defines the
orientation of the complex, and namely this orientation of $D$-simplices
is regarded as positive.

Evidently, interesting for us complex $\mK$ is orientable.

Below index $A$ enumerates 4-simplices.
Introduce the following notation for oriented 1-simplices in the case when
the vertexes $a_i$ and $a_j$ belong to the 4-simlex with index $A$:
\begin{eqnarray}
X^A_{ij}=a_ia_j=-X^A_{ji}\,.
\label{discr5}
\end{eqnarray}

Let
\begin{eqnarray}
s^4_A=a_{i_0}a_{i_1}a_{i_2}a_{i_3}a_{i_4}
\label{discr6}
\end{eqnarray}
be an positively oriented 4-simlex with index $A$.
An oriented frame of a simplex (\ref{discr6}) at a vertex $a_{i_0}$ is
the ordered set of four oriented 1-simplices (\ref{discr5}) such that an
even permutation of these 1-simplices does not change the orientation
while an odd permutation changes the orientation of the frame to the opposite.
By definition, the frame
\begin{eqnarray}
{\cR}^{A\,i_0}=\big(X^A_{i_0i_1},\,X^A_{i_0i_2},\,X^A_{i_0i_3},\,X^A_{i_0i_4}\big)
\label{discr7}
\end{eqnarray}
is oriented positively.

Let $\gamma^a,\,\,a,b,c,\ldots =1,2,3,4$ be $4\times 4$ Dirac
matrices with Euclidean signature. Thus all Dirac matrices as well
matrix
\begin{eqnarray}
\gamma^5=\gamma^1\gamma^2\gamma^3\gamma^4\,, \qquad
\tr\,\gamma^5\gamma^a\gamma^b\gamma^c\gamma^d=4\,\varepsilon
^{abcd}
\label{discr10}
\end{eqnarray}
are Hermitian. To each vertex $a_i$, we assign the Dirac spinors
$\psi_i$ and $\overline{\psi}_i$ each of whose components assumes
values in a complex Grassman algebra. In the case of Euclidean
signature, the spinors $\psi_i$ and $\overline{\psi}_i$ are
independent variables and are interchanged under the Hermitian
conjugation. The Dirac matrixes act from the left to the spinors
$\psi_i$ and from the right to the spinors $\overline{\psi}_i$.

Let us assign to each oriented edge $a_ia_j$ an element of the
group  $Spin(4)$:
\begin{eqnarray}
\Omega_{ij}=\Omega^{-1}_{ji}=\exp\left(\frac{1}{2}\omega^{ab}_{ij}
\sigma^{ab}\right)\,, \ \ \ \sigma^{ab}=\frac{1}{4}[\gamma^a,\,\gamma^b]\,.
\label{discr20}
\end{eqnarray}
Holonomy element $\Omega_{ij}$ of the gravitational field executes
a parallel transformation of spinor $\psi_j$ from vertex $a_j$ of
edge $a_ia_j$ to neighboring vertex $a_i$. We denote by $V$ a
linear space with basis $\gamma^a$. Let each oriented edge
$a_ia_j$ be put in correspondence with element $\hat{e}_{ij}\equiv
e^a_{ij}\gamma^a\in V$, such that
\begin{eqnarray}
\hat{e}_{ij}=-\Omega_{ij}\hat{e}_{ji}\Omega_{ij}^{-1}\,.
\label{discr30}
\end{eqnarray}
The notations
$\overline{\psi}_{Ai}, \ \psi_{Ai}, \ \hat{e}_{Aij}, \
\Omega_{Aij}$ and so on indicate that edge $X^A_{ij}=a_ia_j$ belongs to 4-simplex
with index $A$. Here, the sign "$ \ - \ $" in (\ref{discr30}) is
due to the fact that $e_{A\,ij}$ and $e_{A\,ji}$ are the values of the 1-form on the
edges $X^A_{ij}=a_ia_j$ and $X^A_{ji}=a_ja_i=-a_ia_j=-X^A_{ij}$ (which are oriented mutually oppositely), respectively.
The "facing" from the elements of a holonomy group on the right-hand side of
Eq. (\ref{discr30}) are necessary since the element
$e_{A\,ji}$ must be paralled-translated from the vertex $a_{A\,j}$ to the vertex
 $a_{A\,i}$ to compare this element with the element $e_{A\,ij}$. The quantities
 assigned to each oriented edge $a_ia_j$ and satisfying to Eq. (\ref{discr30})
 are called 1-forms.

By assumption, complex $\mK$ has a disk topology. For such a
complex, the concept of orientation can be introduced. We define
the orientation of the complex by defining the orientation of each
4-simplex. In this case, if two 4-simplices have a common
tetrahedron, the two orientations of the tetrahedron, which are
defined by the orientations of these two 4-simplices, are
opposite. In our case, the complex obviously has only two
orientations.

Let $a_{Ai}, \ a_{Aj}, \ a_{Ak}, \ a_{Al}$, and $a_{Am}$ be all
five vertices of a 4-simplex with index $A$ and
$\varepsilon_{Aijklm}=\pm 1$ depending on whether the order of
vertices $a_{Ai}\,a_{Aj}\,a_{Ak}\,a_{Al}\,a_{Am}$ defines the
positive or negative orientation of this 4-simplex. In addition,
$\varepsilon_{ijklm}=0$ if at least two indices coincide. We can
now write the Euclidean action in the model in question:
\begin{eqnarray}
I=\frac{1}{5\times
24}\sum_A\sum_{i,j,k,l,m}\varepsilon_{Aijklm}\tr\,\gamma^5 \times
\nonumber \\
\times\left\{-\frac{1}{2\,l^2_P}\Omega_{Ami}\Omega_{Aij}\Omega_{Ajm}
\hat{e}_{Amk}\hat{e}_{Aml}+\right.
\nonumber \\
\left.+\frac{1}{24}\hat{\Theta}_{Ami}
\hat{e}_{Amj}\hat{e}_{Amk}\hat{e}_{Aml}\right\}\,,
\label{discr40}
\end{eqnarray}
\begin{gather}
\hat{\Theta}_{Aij}=
\frac{i}{2}\gamma^a\left(\overline{\psi}_{Ai}\gamma^a
\Omega_{Aij}\psi_{Aj}-\overline{\psi}_{Aj}\Omega_{Aji}\gamma^a\psi_{Ai}\right)
\equiv
\nonumber \\
\equiv\Theta^a_{Aij}\gamma^a\in V\,.
\label{discr50}
\end{gather}
The quantity $\hat{\Theta}_{Aij}$ represents an Hermithian operator.
One can easily verify that 1-form (\ref{discr50}), just as the 1-form $\hat{e}_{ij}$, satisfies relation
(\ref{discr30}). This fact is established by the repeated application of the formula
\begin{gather}
S^{-1}\,\gamma^a\,S=S^a_b\,\gamma^b\,,
\label{discr60}
\end{gather}
where
\begin{gather}
S\equiv\exp\frac{1}{2}\,\varepsilon_{ab}\,\sigma^{ab}\,, \qquad \
\varepsilon_{ab}=-\varepsilon_{ba}=\varepsilon^a_b\,,
\nonumber \\
S^a_b\equiv\big(\exp\varepsilon\big)^a_b=
\delta^a_b+\varepsilon^a_b+\frac{1}{2}\,
\varepsilon^a_c\varepsilon^c_b+\,\ldots\,.
\label{discr70}
\end{gather}
It is easy to see that the action (\ref{discr40}) is real.

The volume of a 4-complex is given by
\begin{gather}
V_A=\frac{1}{4!}\times\frac{1}{5!}\times
\nonumber \\
\times\sum_A\sum_{i,j,k,l,m}\varepsilon_{A\,ijklm}\,
\varepsilon^{abcd}\,e^a_{A\,mi}e^b_{A\,mj}e^c_{A\,mk}e^d_{A\,ml}\,.
\label{discr80}
\end{gather}
Here, factor $1/4!$ is required since the volume of a
four-dimensional parallelepiped with generatrices
$e^a_{A\,mi},\;e^b_{A\,mj},\;e^c_{A\,mk}$, and $e^d_{A\,ml}$ is
$4!$ times larger than the volume of a 4-simplex with the same
generatrices, while factor $1/5!$ is due to the fact that all five
vertices of each simplex are taken into account independently.

The dynamic variables are quantities $\Omega_{ij}$ and
$\hat{e}_{ij}$, which describe the gravitational degrees of
freedom, and fields $\overline{\psi}_i$ and $\psi_i$, which are
material fermion fields (other material fields are not considered
here).

In the space of fields, there acts a gauge group according to the
following rule. To each vertex $a_{A\,i}$, let us assign an element of
the group $S_{A\,i}\in\Spin(4)$. According to the principle of gauge
invariance, the fields $\Omega$, \ $e$, \
$\psi$, and the transformed fields
\begin{gather}
\tilde{\Omega}_{A\,ij}=S_{A\,i}\,\Omega_{A\,ij}\,S^{-1}_{A\,j}\,,
\nonumber \\
\tilde{e}_{A\,ij}=S_{A\,i}\,e_{A\,ij}\,S^{-1}_{A\,i}\,,
\nonumber \\
\tilde{\psi}_{A\,i}=S_{A\,i}\,\psi_{A\,i}\,, \ \qquad \
\tilde{\overline{\psi}}_{A\,i}=\overline{\psi}_{A\,i}\,S^{-1}_{A\,i}
\label{discr90}
\end{gather}
are physucally equivalent. This means that the action (\ref{discr40}) is invariant
under the transformations (\ref{discr90}). Under the gauge transformatios
(\ref{discr90}), the 1-form $\Theta$ is transformed in the same way as the form
$e$:
\begin{gather}
\tilde{\hat{\Theta}}_{A\,ij}=S_{A\,i}\,\hat{\Theta}_{A\,ij}\,S^{-1}_{A\,i}\,.
\label{discr100}
\end{gather}
The last formula is verified with the help of Eqs. (\ref{discr60}),
(\ref{discr70}) and (\ref{discr90}). Gauge invariance the action (\ref{discr40})
and the volume (\ref{discr80}) is established by using Eqs. (\ref{discr90}) and (\ref{discr100}).

It is natural to interpret the quantity
\begin{gather}
l^2_{ij}\equiv\frac{1}{4}\,\tr\,(\hat{e}_{ij})^2=\sum_{a=1}^4(e^a_{ij})^2
\label{discr110}
\end{gather}
as the square of the length of the edge $a_ia_j$. Thus, the geometric properties
of a simplicial complex prove to be completely defined.

Now, let us show that, in the limit of slowly varying fields, the action
(\ref{discr40}) reduces to the continuum action of gravity, minimally
connected with with a Dirac field, in a four-dimensional Euclidean space.

Consider a certain subset of vertices from the simplicial complex and assign
the coordinates (real numbers)
\begin{gather}
x^{\mu}_{A\,i}\equiv x^{\mu}(a_{A\,i})\,,
 \qquad \ \mu=1,\,2,\,3,\,4
\label{discr120}
\end{gather}
to each vertex $a_{A\,i}$ from this subset. We stress that these coordinates
are defined only by their vertices rather than by the higher dimension simplices
to whom these vertices belong; moreover, the correspondence between the vertices
from the subset considered and the coordinates (\ref{discr120}) is one-to-one.

Suppose that
\begin{gather}
|\,x^{\mu}_{A\,i}-x^{\mu}_{A\,j}\,|\ll 1\,.
\label{discr130}
\end{gather}
Estimates (\ref{discr130}) can easily be satisfied if the complex contains
a vary large number of simplices and its geometric realization is an almost smooth
four-dimensional surface {\footnote{Here, by an almost smooth surface, we mean a
piecewise smooth surface consisting of flat four-dimensional simplices,
such that the angles between adjacent 4-simplices tend to zero and the sizes of these
simplices are commensurable.}}. Suppose also that the four 4-vectors
\begin{gather}
\d x^{\mu}_{A\,ji}\equiv x^{\mu}_{A\,i}-x^{\mu}_{A\,j}\,,
 \ \ \qquad i\neq j\,, \ \ \ i=1,\,2,\,3,\,4
\label{discr140}
\end{gather}
are linearly independent and
\begin{gather}
\left\vert
\begin{array}{llll}
\d x^1_{A\,m1} \ & \ \d x^2_{A\,m1} \ & \ldots & \ \d x^4_{A\,m1}\\
\ldots  & \ldots  &  \ldots  & \ldots \\
\d x^1_{A\,m4} \ & \ \d x^2_{A\,m4} \ & \ldots & \ \d x^4_{A\,m4}
\end{array}\right\vert >0\,,
\label{discr150}
\end{gather}
provided that the frame $\big(X^A_{m\,1},\,
\ldots\,,\,X^A_{m\,4}\,\big)$ is positively oriented. Inequality
(\ref{discr150}) implies that positively oriented local coordinates
are introduced on the almost flat surface considered. Here, the differentials of coordinates
(\ref{discr140}) correspond to one-dimensional simplices $a_{A\,j}a_{A\,i}$, so that,
if the vertex $a_{A\,j}$ has coordinates $x^{\mu}_{A\,j}$, then the vertex
$a_{A\,i}$ has the coordinates $x^{\mu}_{A\,j}+\d x^{\mu}_{A\,ji}$.

In the continuum limit, the holonomy group elements (\ref{discr20}) are
close to the identity element, so that the quantities $\omega^{ab}_{ij}$
tend to zero being of the order of $O(\d x^{\mu})$.
Thus one can consider the following system of equation for $\omega_{A\,m\mu}$:
\begin{gather}
\omega_{A\,m\mu}\,\d x^{\mu}_{A\,mi}=\omega_{A\,mi}\,,  \ \ \
i=1,\,2,\,3,\,4\,.
\label{discr160}
\end{gather}
In this system of linear equation, the indices $A$ and $m$ are
fixed, the summation is carried out over the index $\mu$, and
index runs over all its values. Since the determinant
(\ref{discr150}) is positive, the quantities $\omega_{A\,m\mu}$
are defined uniquely. Suppose that a one-dimensional simplex
$X^A_{m\,i}$ belong to four-dimensional simplices with indices
$A_1,\,A_2,\,\ldots\,,\,A_r$. Introduce the quantity
\begin{gather}
\omega_{\mu}\left[\frac{1}{2}\,(x_{A\,m}+
x_{A\,i})\,\right]\equiv\frac{1}{r}\,
\bigg\{\omega_{A_1\,m\mu}+\,\ldots\,+\omega_{A_r\,m\mu}\,\bigg\}\,,
\label{discr170}
\end{gather}
which is assumed to be related to the midpoint of the segment
$[x^{\mu}_{A\,m},\,x^{\mu}_{A\,i}\,]$. Recall that the coordinates
$x^{\mu}_{A\,i}$ just as the differentials (\ref{discr140}),
depend only on vertices but not on the higher dimensional
simplices to which these vertices belong. According to the
definition, we have the following chain of equalities:
\begin{gather}
\omega_{A_1\,mi}=\omega_{A_2\,mi}= \,\ldots\,=\omega_{A_r\,mi}\,.
\label{discr180}
\end{gather}
It follows from (\ref{discr140}) and
(\ref{discr160})--(\ref{discr180}) that
\begin{gather}
\omega_{\mu}\left(x_{A\,m}+ \frac{1}{2}\,\d x_{A\,mi}\,\right)\,\d
x^{\mu}_{A\,mi}= \omega_{A\,mi}  \,.
\label{discr190}
\end{gather}
The value of the field $\omega_{\mu}$ in (\ref{discr190}) on each
one-dimensional simplex is uniquely defined by this simplex.

Next, we assume that the fields $\omega_{\mu}$ smoothly depend on
the points belonging to the geometric realization of each
four-dimensional simplex. In this case, the following formula is
valid up to $O\big((\d x)^2\big)$ inclusive:
\begin{gather}
\Omega_{A\,mi}\,\Omega_{A\,ij}\,\Omega_{A\,jm}=
\exp\left[\frac{1}{2}\,\mR_{A\,m\mu\nu}\,\d x^{\mu}_{A\,mi}\, \d
x^{\nu}_{A\,mj}\,\right]\,,
\label{discr200}
\end{gather}
where
\begin{gather}
\mR_{A\,m\mu\nu}=\partial_{\mu}\omega_{A\,m\nu}-\partial_{\nu}\omega_{A\,m\mu}+
[\omega_{A\,m\mu},\,\omega_{A\,m\nu}\,]\,.
\label{discr210}
\end{gather}
On the right-hand side of (\ref{discr200}), as well as in equality
(\ref{discr210}), all fields are taken at the vertex $a_{A\,m}$ of
a four-dimensional simplex $A$ as is indicated by the subscript
$A\,m$. When deriving formula (\ref{discr200}), we used the
Hausdorff formula.

In exact analogy with (\ref{discr160}), let us write out the following relations
for a tetrad field without explanations:
\begin{gather}
e_{A\,m\mu}\,\d x^{\mu}_{A\,mi}=e_{A\,mi}\,.
\label{discr220}
\end{gather}

Using (\ref{discr20}) and (\ref{discr160}), we can rewrite the 1-form
(\ref{discr50}) as
\begin{gather}
\Theta_{A\,ij}=\gamma^a\,\frac{i}{2}\,\left[\,
\overline{\psi}_{A\,i}\,\gamma^a\,{\cal D}_{\mu}\,\psi_{A\,i}-
\overline{{\cal D}_{\mu}\,\psi}_{A\,i}\,\gamma^a\,\psi_{A\,i}\,
\right]\,\d x^{\mu}_{A\,ij}\,,
\label{discr230}
\end{gather}
to within $O(\d x)$; here,
\begin{gather}
{\cal D}_{\mu}\,\psi_{A\,i}=\partial_{\mu}\,
\psi_{A\,i}+\omega_{A\,i\mu}\,\psi_{A\,i}\,.
\label{discr240}
\end{gather}

Before rewriting the action (\ref{discr40}) in the continuum limit,
we give the following obvious formula:
\begin{gather}
\sum_{\sigma(Am)}\,\varepsilon_{\sigma(Am)}\,
\d x^{\mu}_{A\,mi}\,\d x^{\nu}_{A\,mj}\,\d x^{\lambda}_{A\,mk}\,
\d x^{\rho}_{A\,ml}=
\nonumber \\
=24\,\varepsilon^{\mu\nu\lambda\rho}\,
v_{S\,A}\,.
\label{discr250}
\end{gather}
Here, $\varepsilon^{\mu\nu\lambda\rho}$ is a completely antisymmetric symbol,
which is equal to unity when $(\mu\,\nu\,\lambda\,\rho)=(1\,2\,3\,4)$ (compare
with (\ref{discr150})), and $v_{S\,A}$ is the volume of the geometric realization
of simplex $A$ in a four-dimensional Euclidean space when the Euclidean coordinates
of the geometric realization of the simplex are equal to the corresponding
coordinates of its vertices (\ref{discr120}). The factor 24 in (\ref{discr250})
is necessary since the volume $v_{S\,A}$ of the four-dimensional simplex
on the right-hand side is less than the volume of a four-dimensional parallelepiped
constructed on the vectors $\d x^{\mu}_{A\,mi},\,\ldots\,,\,\d x^{\mu}_{A\,ml}$
by a factor of 24.

Applying formulas (\ref{discr200})--(\ref{discr250}) and changing the
summation to integration, we obtain the following expression for the
action (\ref{discr40}) in the continuum limit:
\begin{gather}
I=\int\,\tr \gamma^5\times
\nonumber \\
\times\left[-\frac{1}{4\,l^2_P}\,
\left(\mR+\frac{1}{3}\,\Lambda\,e\wedge e\,\right)+
\frac{1}{24}\,\Theta\wedge e\right]\wedge e\wedge e\,.
\label{discr260}
\end{gather}
Here, the curvature 2-form (see (\ref{discr210})) and the 1-forms
(see (\ref{discr220}), (\ref{discr230})) are defined by
\begin{gather}
\mR\equiv\frac{1}{2}\,\sigma^{ab}\,
R^{ab}_{\mu\nu}\,\d x^{\mu}\wedge\d x^{\nu}\,,
\nonumber \\
e=\gamma^a\,e^a_{\mu}\,\d x^{\mu}\,,
\nonumber \\
\Theta=\gamma^a\,\frac{i}{2}\left[\,\overline{\psi}
\gamma^a\,{\cal D}_{\mu}\,\psi-\overline{{\cal D}_{\mu}\psi}
\,\gamma^a\,\psi\,\right]\,\d x^{\mu}\,.
\label{discr270}
\end{gather}

Thus, in the continuum limit, the action
(\ref{discr40}) proves to be equal to the action of gravity with a
$\Lambda$-term and a metric with Euclidean signature that is minimally
connected with a Dirac field.

\subsection{Quantization of Discrete Gravity}

Let us determine the partition function $Z$ for a discrete
Euclidean gravity, which becomes the transfer-matrix in discrete quantum gravity
after passing to the Lorentzian signature. Let us enumerate the
zero-dimensional (vertices) and one-dimensional (edges) simplices
by indices $\cV$ and $\cE$, respectively, and denote by
$\psi_{\cV}$, \ $\Omega_{\cE}$, rtc. the corresponding variables.
By definition,
\begin{gather}
Z=\const\cdot\bigg (\prod_{\cE}\,\int\,
\d\Omega_{\cE}\,\int\,\d e_{\cE}\,\bigg)\times
\nonumber \\
\times\big(\prod_{\cV}\,\d\overline{\psi}_{\cV}\,
\d\psi_{\cV}\,\big)\,\exp\big(-I\,\big)\,.
\label{discr280}
\end{gather}
Here, $\const$ is a certain normalizing factor, $\d\Omega_{\cE}$ is the
Haar measure on the group $\Spin(4)$,
\begin{gather}
\d e_{\cE}\equiv\prod_a\,\d\omega^a_{\cE}\,,
\qquad \ e_{\cE}=\omega^a_{\cE}\,\gamma^a\,,
\label{discr290}
\end{gather}
and
\begin{gather}
\d\overline{\psi}_{\cV}\,\d\psi_{\cV}\equiv\prod_{\nu}\,
\d\overline{\psi}_{\cV\nu}\,\d\psi_{\cV\nu}\,.
\label{discr300}
\end{gather}
The index $\nu$ in (\ref{discr300}) enumerates individual components of the
spinors $\psi_{\cV}$ and $\overline{\psi}_{\cV}$, such that we have
a product of the differentials of all independent generators of the
Grassman algebra of Dirac spinors in (\ref{discr300}). The action $I$ in
(\ref{discr280}) is defined by formula
(\ref{discr40}).

Note that the measure (\ref{discr290}) is determined correctly in view
of invariance of the Haar measure and the relations
(\ref{discr20}) and (\ref{discr30}). Therefore, one can really assume
that the measure (\ref{discr290}) is related to the set of edges.

Obviously, all the measures used in the functional integral
(\ref{discr280}) are invariant under the gauge transformations (\ref{discr90}).
Since the action $I$ (\ref{discr40}) in (\ref{discr280}) is also
gauge invariant, the partition function (\ref{discr280}) is invariant
under the action of the gauge group (\ref{discr90}).

Consider the partition function (\ref{discr280}) with a zero
$\Lambda$-term in the absence of fermions. In this case, the
integral over the 1-form $e_{\cE}$ becomes Gaussian:
\begin{gather}
Y\big\{\Omega\,\big\}=\int\,D\,z\cdot\exp
\left(\frac{1}{2}\,z_m\,{\ccm}_{m\,n}\,z_n\right)\,.
\label{discr310}
\end{gather}
Here, $\{\,z_m\,\}$, \ $m=1,\,\ldots\,,\,Q$ denotes a set of real
variables $\{\omega^a_{\cE}\}$ and ${\ccm}_{mn}$ is a real symmetrical
matrix depending on the elements of the holonomy group $\Omega_{\cE}$.
Thus,
\begin{gather}
\frac{1}{2}\,z_m\,{\ccm}_{mn}\,z_n\equiv
\frac{1}{l^2_P}\,\frac{1}{5}\cdot\frac{1}{24}\,\sum_{A,\,m}\,
\sum_{\sigma(A\,m)}\,\varepsilon_{\sigma(Am)}\times
\nonumber \\
\times\tr\big(\gamma^5\,\Omega_{A\,mi}\,
\Omega_{A\,ij}\,\Omega_{A\,jm}\,e_{A\,mk}\,e_{A\,ml}\,\big)\,.
\label{discr320}
\end{gather}

Denote by $\{\,\lambda_q\,\}$, where $q=1,\,\ldots\,,\,Q$,
a set of eigenvalues of the matrix ${\ccm}_{mn}$. Let
$\varepsilon_q=\sign\lambda_q$. Since, in general, there are both
negative and positive eigenvalues among $\{\lambda_q\}$, the integral
(\ref{discr310}) should be redefined. This is done by passing to Lorentzian
signature. Under this procedure, the eigenvalues are transformed by
the rule
$$
\lambda_q\rightarrow e^{i\varphi}\,\lambda_q\,,
$$
where $\varphi=0$ in the Euclidean space and $\varphi=\pi/2$
in the case of the Minkowski signature. Thus, the Euclidean
Gaussian integral
\begin{gather}
{\cI}_E=\frac{1}{\sqrt{2\pi}}\,\int_{-\infty}^{+\infty}\,
\d z\cdot\exp\left(\frac{1}{2}\,\lambda\,z^2\,\right)
\label{discr330}
\end{gather}
reduces to the Fresnel integral in the Minkowski signature:
\begin{gather}
{\cI}_M=\frac{1}{\sqrt{2\pi}}\,\int_{-\infty}^{+\infty}\,
\d z\cdot\exp\left(\frac{i}{2}\,\lambda\,z^2\,\right)=
\sqrt{\frac{i}{\lambda}}=
\nonumber \\
=(i)^{\frac{\varepsilon}{2}}\,
\frac{1}{\sqrt{|\,\lambda\,|}} \,,
\label{discr340}
\end{gather}
where $\varepsilon=\sign\lambda$. Let us perform the analytic continuation
$$
\lambda\rightarrow
e^{-i\varphi}\,\lambda
$$
on the right-hand side of Eq. (\ref{discr340}) and set $\varphi=\pi/2$.
Thus, we recover the Euclidean signature of a metric and
obtain the following value for integral (\ref{discr330}):
\begin{gather}
{\cI}_E=(i)^{\frac{\varepsilon+1}{2}}\,
\frac{1}{\sqrt{|\,\lambda\,|}}\,.
\label{discr350}
\end{gather}

Now, using Eq. (\ref{discr350}), we redefine the integral
(\ref{discr310}) of interest:
\begin{gather}
Y\big\{\Omega\,\big\}=\const\,\prod_q\,
(i)^{\frac{\varepsilon_q+1}{2}}\,|\,\lambda_q\,|^{-1/2}\,.
\label{discr360}
\end{gather}

If there are fermion fields in the theory, one should first
calculate a functional integral over fermions. The subsequent
integration over the 1-form $e$ remains Gaussian and yields a
contribution of the form (\ref{discr360}) to the partition
function. The remaining integral over the elements of the holonomy
group $\Omega$ may prove to be divergent despite the compactness
of this group. Indeed, certain eigenvalues $\lambda_q$ may vanish
under certain configurations of the field $\Omega$. Since the
expression under the integral sign depends on the negative powers
of $\lambda_q$, the integral over the field $\Omega$ may prove to
be divergent. From the physical point of view, these divergences
are of great interest. Note that the tendency of eigenvalues
$\lambda_q$ to zero implies that the integral over the 1-form
$e^a$ is saturated when the absolute values of this field $e^a$
(or its certain components) tend to infinity. This means that the
size of universe tends to infinity (see (\ref{discr110})). On the
other hand, as will be shown below, the fact that the field
components $e^a$ have large values implies that the dynamics of
the system becomes quasiclassical. Therefore, from the physical
viewpoint, these divergences imply birth of quasiclassical
macroscopic space-time.

Concerning the problem under discussion, we note that the presence
of Dirac fields in integral (\ref{discr280}) only strengthens the
divergence under the integration over the field $e^a$. Indeed,
after the integration over the fermion field, the integral over
the field $e^a$ is rewritten as (cf. (\ref{discr330}) and
(\ref{discr340}))
\begin{gather}
{\cI}=\frac{1}{\sqrt{2\pi}}\,\int_{-\infty}^{+\infty}\, \d
z\,P_n(z)\cdot\exp\left(\frac{i}{2}\,\lambda z^2\,\right)\,,
 \label{discr370}
\end{gather}
where $P_n(z)$ is a polynomial in $z$ of degree $n$. For small
$\lambda$, integral (\ref{discr370}) is proportional to
$|\,\lambda\,|^{-(n+1)/2}$.

A similar physical interpretation of divergences under the
integration over the field $e^a$ in the continuum quantum
$B$-$F$-theory in a three-dimensional space-time was given by
Witten in \cite{15}.

Let us notice another possible type of divergences in a discrete
quantum gravity. If the partition function (\ref{discr280}) was
defined for a metric with Lorentzian signature, then the elements
of the holonomy group would be the noncompact group
$\Spin(3,\,1)$. The gauge group (\ref{discr90}) would also be
noncompact, being a direct product of the $\cV$ copies of the
group $\Spin(3,\,1)$. Since both the measure and action in the
transfer-matrix are gauge invariant, the functional integral in
the transfer-matrix would not be defined at all before the
fixation (at least partial) of the gauge. However, the fixation of
the gauge in the fundamental transfer-matrix seems to be a so
artificial procedure that the theory itself looses its beauty and
sense. In our opinion, this means that the fundamental partition
function for a discrete theory of gravity can be constructed only
on the basis of a metric with Euclidean signature.

In their well-known paper \cite{16}, Hartle and Hawking made a
hypothesis that the wave function of the universe must be
calculated with the use of the functional integral on the basis of
a metric with Euclidean signature. But in the case of the gravity
theory the Euclidean action is not positively defined. In our
opinion, the arguments for a metric with Euclidean signature
provided by the discrete theory of gravity are much more reliable
than the arguments given in \cite{16}.

\subsection{High Temperature Expansion}

From the beginning let us consider the integral (\ref{discr280}) in the
region of integration variables where
\begin{gather}
|e_{ij}^a|>l_0\gg l_P\,. \label{discr380}
\end{gather}
In this region each item in the sum (\ref{discr40}) generally is also
large since the items in the sum (\ref{discr40}) are polynomials in the variables
$e_{ij}^a$ of powers not less than two. Therefore the whole integral in
(\ref{discr280}) can be estimated quasiclassically ore by the stationary phase
method. In this region one must use the long wave limit action
(\ref{discr260}), and to perform the stationary phase calculations the integration paths
in (\ref{discr280}) must be deformed so that Lorentzian signature is realized.
Thus the time arises.
We see that in considered model the time arises dynamically in continual limit.
The study of continual limit of the theory is performed in
the subsequent sections.

Now let us consider the integral (\ref{discr280}) in the
region of integration variables where
\begin{gather}
|e_{ij}^a|<l_1\ll l_P\,.
\label{discr390}
\end{gather}
In this region each item in the sum (\ref{discr40}) is small, so that
the subintegral quantity in (\ref{discr280}) (in the case of pure gravity
and zero $\Lambda$-term) can be written as
\begin{gather}
\exp\big(-I\,\big)=\prod_{A}\prod_{i,j,k,l,m}
\bigg(1+\frac{1}{5\times24\times l^2_P}\,\varepsilon_{Aijklm}\tr\,\gamma^5 \times
\nonumber \\
\times\Omega_{Ami}\Omega_{Aij}\Omega_{Ajm}
\hat{e}_{Amk}\hat{e}_{Aml}\bigg)\,.
\label{discr400}
\end{gather}
The expansion (\ref{discr400}) is called further as high temperature expansion.
It is well known that the analogous representation for the $\exp\big(-I\,\big)$
is true in the lattice Yang--Mills theory in the limit of large coupling constant.
From such representation the significant phenomenon of colour confinement follows.
Originally the phenomenon of colour confinement has been obtained analytically with
the help of
high temperature expansion (with the help of representation of the tipe
(\ref{discr400})) by Wilson, and then numerous computer simulations
confirmed this conclusion. Since the situations concerning
high temperature expansion in both theories are closely analogous,
we make the conclusion that in the region of variables
(\ref{discr390}) also take place colour confinement. Introduce the following
notations:
$$
C=\{a_{i_0}a_{i_1},\,a_{i_1}a_{i_2},\ldots\,,a_{i_r}a_{i_0}\}
$$
is a closed contour ore a one dimensional subcomplex with zero boundary;
$$
W\,(C)=\langle\,\tr\left(\Omega_{i_0i_1}\Omega_{i_1i_2}\ldots \Omega_{i_ri_0}\right)\rangle_1
$$
is Wilson loop correlator which in our case is calculated in
the theory of pure gravity with zero $\Lambda$-term in the region
of variables restricted by inequalities (\ref{discr390}); $\sigma_C$ is a
two dimensional subcomplex with boundary $\partial\,\sigma=C$;
$n_C(\sigma)$ is the number of triangles containing in $\sigma$ and
$$
n_C=\min_{\sigma}\{n_C(\sigma)\}\,.
$$
Then the simple calculations give the following estimation:
\begin{gather}
W\,(C)\sim\exp\left(-n_C\mu\,\ln l_1^{-1}\right)\,.
\label{discr410}
\end{gather}
Here $\mu$ is a number which does not depend on contour $C$ and parameter $l_1$.

Let us emphasize that in the case of discrete quantum gravity the role
of colour gauge group play the group (\ref{discr90}). Thus only singlet
(i.e. scalar, but not spinor, vector and so on) fields
with respect to the group (\ref{discr90}) have quasiparticle excitations
in the region (\ref{discr390}), i.e. on the early stages of universe development.
This conclusion partially justifies the use only scalar fields in
numerous works in which the dynamics of early universe is investigated.
But in contrast to the Yang--Mills theory in expanding universe
 the phase transition occurs to
deconfinement phase (formally in the region (\ref{discr380})). In
this phase the dynamics becomes quasiclassical.

Let us now show that the modes of quantized fields in the
quasiclassical continual phase have essentially noncompact packing
in momentum space. This important conclusion follows from high
temperature expansion and the most general properties of spectrum
of elliptic operators.

We illustrate the effect in Appendix A on the example of the
spectrum of one dimensional discrete Laplace operator on random
lattice on a cycle. In the cases of 3 and 4 vertexes the problem
is solved exactly and we see that in the case when the total
length of the cycle is fixed but the distances between some
vertexes tend to zero some of eigenfunctions of the operator tend
to infinity as inverse degrees of the small distances between the
corresponding vertexes.

Let us make the estimation of modes packing in our theory in 3-dimensional
space. We keep in mind the scalar field since the spinor structure
does not affect significantly for the estimation.

Firstly, we write out the trivial formula for for the volume in momentum
space occupied by $N$ modes placed in the flat volume $V$ and densely packed
in momentum space:
\begin{gather}
\Omega=(2\pi)^3\frac{N}{V}\,.
\label{discr420}
\end{gather}

Now, one must take into account the fact that in confinement phase
all correlators of fundamental fields drop exponentially with space separation.
This means that the fields at nearest regions of space volume are not correlated.
The same conclusion remains true at initial times in quasiclassical phase.
Therefore let us divide a macroscopic volume $V$ with the total number of
degrees of freedom (ore the number of modes) $N$ into ${\cal N}$ subvolumes
$v_i$ in each of which contains $n_i$ degrees of freedom. Thus
\begin{gather}
\sum_{i=1}^{{\cal N}}n_i=N\,, \qquad  \sum_{i=1}^{{\cal N}}v_i=V\,,
\label{discr430}
\end{gather}
and
\begin{gather}
\omega_i=(2\pi)^3\frac{n_i}{v_i}
\label{discr440}
\end{gather}
is the minimal possible volume in momentum
space occupied by $n_i$ modes placed in the flat volume $v_i$.
Now instead of the quantity (\ref{discr420}) one must consider the
following quantity:
\begin{gather}
\tilde{\Omega}=\frac{(2\pi)^3}{{\cal N}}\sum_{i=1}^{{\cal N}}\frac{n_i}{v_i}\,.
\label{discr450}
\end{gather}
Indeed, the minimum of quantity (\ref{discr450}) subjected to the
constraints (\ref{discr430}) is equal to (\ref{discr420}).

But in considered theory the volumes $v_i$ are variable quantities.
Therefore one must introduce the measure on the manifold of volumes $\{v_i\}$.
The simplest measure agreed with fundamental measure (\ref{discr290}) looks like
the following:
\begin{gather}
\d\mu=\frac{({\cal N}-1)!}{V^{{\cal N}-1}}\delta\left(V-\sum_{i=1}^{{\cal N}}v_i\right)
\prod_{i=1}^{{\cal N}}\d v_i\,, \quad v_i>0\,,
\nonumber \\
\int\d\mu=1\,.
\label{discr460}
\end{gather}
Hence instead of (\ref{discr450}) the more physically sensible quantity is
\begin{gather}
\langle\tilde{\Omega}\rangle\equiv\int\tilde{\Omega}\d\mu=(2\pi)^3\frac{{\cal
N}-1}{V\,{\cal N}}\sum_{i=1}^{{\cal N}}n_i\int_{v_i\ll V}\frac{\d
v_i}{v_i}=
\nonumber \\
=(2\pi)^3\frac{N}{V}\int_{v_i\ll V}\frac{\d v_i}{v_i}\,.
 \label{discr470}
\end{gather}
The last equality is obtained taking into account the first
constraint of (\ref{discr430}) and the relation ${\cal N}\gg 1$.

The comparison of Eqs. (\ref{discr420}) and (\ref{discr470}) shows
that taking into account the dynamics of the system leads to the
essential expansion of the momentum space volume occupied by
quantum field modes. This expansion factor is
\begin{gather}
\varkappa_1=\int_{v_i\ll V}\frac{\d
v_i}{v_i}=3\ln\frac{a_1}{a_0}=3\ln\xi_0\,.
\label{discr480}
\end{gather}
Here $a_0$ is some minimal dimension of the theory and $a_1\sim
V^{1/3}$.

Now there is a need to make a kind of renorm-group. Let $n$ be the
number of steps of renorm-group and
\begin{gather}
\xi_s=\frac{a_{s+1}}{a_s}=\xi\gg 1\,, \quad s=1,\ldots\,,n\,,
\label{discr490}
\end{gather}
and $a_{n+1}=a$ is the radius of universe. Thus $\xi^n=a/a_0$. Let
us take
\begin{gather}
n=\frac{1}{\lambda}\ln\frac{a}{a_0}\gg 1\,, \quad \lambda\gg 1\,.
\label{discr500}
\end{gather}
Using Eqs. (\ref{discr480})--(\ref{discr500}) it is easy to see
that the expansion factor of momentum space volume occupied by
modes after $n$ steps is
\begin{gather}
\varkappa_n=\prod_{s=1}^n(3\ln\xi_s)=(3\ln\xi)^n=\left(\frac{a}{a_0}\right)^{(\ln
3\lambda)/\lambda}\,.
\label{discr510}
\end{gather}
The value of right hand side of Eq. (\ref{discr510}) can bee very
large (many orders) in magnitude.

It follows from the presented analysis, that the continual quantum
gravity arising from the discrete quantum gravity (if it exists)
possess very unusual properties. The description of possible such
theory and some consequence is performed in the subsequent
sections.

\section{Method of Dynamic Quantization }

Specific results of the application of the Dynamic quantization
method to the two-dimensional theories \cite{4, 17}, obtained by
explicit constructions and direct calculations, justify the
abstract assumptions and axioms on which this method is based.

We shall explain the ideology and logical scheme of the Dynamic
method taking account of the experience in quantizing
two-dimensional gravity.

The key point in the quantization of two-dimensional gravity was
the construction of a complete set of such operators $ \{A_n,
B_n,\ldots \} $, designated below as $\{A_N, \,A_N^{\dag}\}$, which
possess the following properties:

\centerline{}

1) The operators $A_N$ and $A_N^{\dag}$ are Hermithian conjugates
of one another and
\begin{gather}
[\,A_N,\,A_M\,]=0, \qquad [A_N,\,A^{\dag}_M\,]=\delta_{NM}\,.
\label{dq11}
\end{gather}

2) The set of operators $\{A_N,\,A_N^{\dag}\}$ describes all
physical dynamical degrees of freedom of a system.

3) Each operator from the set $\{A_N,\,A_N^{\dag}\}$ commutes
weakly with all first class constraints or with the complete
Hamiltonian of the theory.

\centerline{}

Quantization is performed directly using the operators
$\{A_N,\,A_N^{\dag}\}$. It means that the space of physical states
is created using the operators $\{A_N^{\dag}\}$ from the ground
state and all operators are expressed in terms of the operators
$\{A_N,\, A_N^{\dag}\}$, as well as in terms of the operators
describing the gauge degrees of freedom. However, in the theory of
two-dimensional gravity the operators
 $\{A_N,\,A_N^{\dag}\}$ were constructed explicitly (i.e., they were
 expressed explicitly in terms of the fundamental dynamical variables),
in more realistic theories this problem is hardly solvable.
Therefore, the set of operators $\{A_N,\,A_N^{\dag}\}$ with
properties 1)--3) must be introduced axiomatically. Conversely, the
properties 1)--3) make it possible, in principle, to express the
initial variables in terms of the convenient operators
$\{A_N,\,A_N^{\dag}\}$.

However, in contrast to the two-dimensional theory of gravity,
regularization is necessary in real models of gravity. In the
Dynamic quantization method, regularization is carried out
precisely in terms of the operators $\{A_N,\,A_N^{\dag}\}$. As will
be shown below, such regularization is natural in generally
covariant theories, since it preserves the form of the Heisenberg
equations and thereby also the general covariance of the theory.

As a result we have the regularized general covariant theory
describing quantum gravity, the main property of which is the
finiteness of physical degrees of freedom contained in each finite
volume. Moreover, the packing
of field modes in momentum space can be made rare.
Evidently, the theory of discrete quantum gravity described
in Section 2 possess the same properties.
Therefore, one can
think that the theory of gravity quantized by dynamic quantization
method is the continuous limit of discrete quantum gravity.

Let's consider a generally covariant field theory. Let us assume
that in this theory the Hamiltonian in the classical limit is an
arbitrary linear combination of the first class constraints and
there are no the second class constraints.

Let $\{\Phi^{(i)}(x),\,P^{(i)}(x)\}$ be a complete set of
fundamental fields of the theory and their canonically conjugate
momenta, in terms of which all other physical quantities and
fields of the theory are expressed. Here the index $(i)$
enumerates the tipe of fields. For example, for some $(i)$ these
can be either 6 spatial components of the metric tensor
$g_{ij}(x)$ or the scalar field $\phi(x)$ or the Dirac field
$\psi(x)$ etc. The set of fields $\{\Phi^{(i)}(x)\}$ is a complete
set of the mutually commuting (at least in formal unregularized
theory) fundamental fields of the theory.

Next, to simplify the notation the index $i$ will be omitted. It
can be assumed that the variable $x$ includes, besides the spatial
coordinates, the index $i$ also.

The construction of a quantum theory by the Dynamic method is
based on the following natural assumptions relative to the
structure of the space $F$ of the physical states of
the theory. \\

{\bf {1.}} {\it {All states of the theory having physical sense
are obtained from the ground state}} $\vert \,0\,\rangle$ {\it
{using the creation operators}} $A^{\dag}_N$:
 \begin{gather}
 \vert \,n_1,\,N_1;\ldots
 ;\,n_s,\,N_s\,\rangle=
 \nonumber \\
 =(n_1!\cdot\ldots\cdot n_s!\,)^{-\frac{1}{2}} \cdot
 (A^{\dag}_{N_1})^{n_1}\cdot\ldots\cdot
 (A^{\dag}_{N_s})^{n_s}\,\vert \,0\,\rangle \ ,
\nonumber \\
   A_N\,\vert\,0\,\rangle=0 \,.
 \label{dq12}
\end{gather}
{\it {States (\ref{dq12}) form an orthonormal basis of the space $F$ of
physical states of the theory.}} \\

The numbers $n_1,\ldots,n_s$ assume integer values and are called
occupation numbers. \\

{\bf {2.}} {\it {The set of states}}
  $\Phi(x)\,\vert\,n_1,N_1;\ldots;n_s,N_s\,\rangle$,
  {\it {, where the set of numbers $(n_1, N_1; \ldots; n_s, N_s)$ is
fixed, contains a superposition of {\bf {all}} states of the
theory, for which one of the occupation number differs in modulus
by one and all other occupation numbers equal to the occupation
numbers of state (\ref{dq12}).}} \\

Here the operators $A^{\dag}_N$ and their conjugates $A$ are
the generators of Heisenberg algebra. The operators
$\{A_N,\,A_N^{\dag}\}$, generally speaking, can be bosonic or
fermionic. If the creation and annihilation operators follow the
Fermi statistics, then the anticommutator are used. For
the case of compact spaces which is interesting for us, we can
assume without loss of generality that the index $N$, enumerating
the creation and annihilation operators, belongs to a discrete
finite-dimensional lattice. A norm can be easily introduced in the
space of indexes $N$.

Since states (\ref{dq12}) are physical, they satisfy the relations
\begin{gather}
{\cal H}_T \, \vert \,n_1,\,N_1;\ldots;\,n_s,\,N_s\,\rangle=0\,\,,
\label{dq13}
\end{gather}
where ${\cal H}_T$ is the complete Hamiltonian of the theory. We
assume that ${\cal H}_T=\sum_{\Xi}v_{\Xi}\chi_{\Xi}$, where
$\{\chi_{\Xi}\}$ is the complete set of the first class
constraints and $\{v_{\Xi}\}$ is arbitrary set of Lagrange
multipliers.

Equations (\ref{dq12}) and (\ref{dq13}) are compatible if and only if the
following relations are valid:
\begin{gather}
[A_N,\,\chi_{\Xi}]=\sum_{\Pi}c_{N\Xi\Pi}\,\chi_{\Pi}\,,
\nonumber \\
[A_N^{\dag},\,\chi_{\Xi}]=-\sum_{\Pi}\chi_{\Pi}\,c_{N\Xi\Pi}^{\dag}=
\sum_{\Pi}\tilde{c}_{N\Xi\Pi}\,\chi_{\Pi}\,.
\label{dq14}
\end{gather}
Since the coefficients $c_{N\Xi\Pi}, \; \tilde{c}_{N\Xi\Pi}$ in
Eq. (\ref{dq14}) generally are operators, the arrangement of the
multiplies in the right hand sides of Eqs. (\ref{dq14}) is important.

Let $(A^{\dag}_N,\,A_N)$ be a pair of bose or fermi creation and
annihilation operators creating or annihilating the state with the
wave function $\psi_N(x)$. According to (\ref{dq14}) we have:
\begin{gather}
[A_N,\,{\cal
H}_T]=\sum_{\Xi,\,\Pi}r_{N\,\Xi\,\Pi}v_{\Xi}\,\chi_{\Pi}
\longleftrightarrow
\nonumber \\
\longleftrightarrow [A_N^{\dag},\,{\cal
H}_T]=-\sum_{\Xi,\,\Pi}\chi_{\Pi}\,v_{\Xi}^*r_{N\,\Xi\,\Pi}^{\dag}\,.
\label{dq15}
\end{gather}
Let an arbitrary operator $\Phi$ be represented as a normal
ordered power series in operators $(A^{\dag}_N,\,A_N)$:
\begin{gather}
\Phi=\Phi'+\phi_NA_N+A^{\dag}_N\tilde{\phi}_N\,.
\label{dq16}
\end{gather}
By definition, here the operator $\Phi'$ does not depend on the
operators $(A^{\dag}_N,\,A_N)$:
\begin{gather}
[\Phi',\,A^{\dag}_N]=[\Phi',\,A_N]=0\,.
\label{dq17}
\end{gather}
It follows from Eqs. (\ref{dq15})--(\ref{dq17}) that
\begin{gather}
[\Phi,\,{\cal H}_T]=[\Phi',\,{\cal
H}'_T]+
\nonumber \\
+\sum_{\Xi}(q_{\Xi}\chi_{\Xi}+\chi_{\Xi}\tilde{q}_{\Xi})+(p_NA_N+A_N^{\dag}\tilde{p}_N)\,.
\label{dq18}
\end{gather}
Here the total Hamiltonian ${\cal H}_T$ is represented according
to (\ref{dq16}), so that ${\cal H}'_T$ does not depend on the operators
$(A^{\dag}_N,\,A_N)$. To verify Eq. (\ref{dq18}) let's write out the
following chain of equalities:
\begin{gather}
[\Phi,\,{\cal H}_T]=[\Phi',\,{\cal H}_T]+\left(\phi_N[A_N,\,{\cal
H}_T]+[A_N^{\dag},\,{\cal H}_T]\,\tilde{\phi}_N\right)+
\nonumber \\
+\left([\phi_N,\,{\cal H}_T]\,A_N+A_N^{\dag}[\tilde{\phi}_N,\,{\cal
H}_T]\right)\,.
\label{dq19}
\end{gather}
As a consequence of Eqs. (\ref{dq15}) the second item in the right hand
side of Eq. (\ref{dq19}) has the same structure as the second item in the
right hand side of Eq. (\ref{dq18}). Evidently, the last items in the
right hand side of Eq. (\ref{dq19}) has the same structure as the last
items in the right hand side of Eq. (\ref{dq18}). Now let's write out the
following identity:
\begin{gather}
[\Phi',\,{\cal H}_T]\equiv[\Phi',\,{\cal H}'_T]+[\Phi',\,{\cal
H}_T-{\cal H}'_T]\,.
\label{dq20}
\end{gather}
By definition
\begin{gather}
{\cal H}_T-{\cal H}'_T=h_N\,A_N+A^{\dag}_N\,\tilde{h}_N\,.
\label{dq21}
\end{gather}
It follows from (\ref{dq17}) and (\ref{dq21}) that
\begin{gather}
[\Phi',\,{\cal H}_T-{\cal
H}'_T]=[\Phi',\,h_N]\,A_N+A^{\dag}_N\,[\Phi',\,\tilde{h}_N]\,.
\label{dq22}
\end{gather}
Combining Eqs. (\ref{dq19}), (\ref{dq20}) and (\ref{dq22}) we come to the Eq. (\ref{dq18}).

Now let's impose an additional pair of second class constraints
\begin{gather}
A_N=0\,, \qquad A^{\dag}_N=0\,.
\label{dq23}
\end{gather}
By definition under the constraints (\ref{dq23}) any operator $\Phi$ is
reduced to the operator $\Phi'$ in (\ref{dq16}). The Dirac bracket arising
under the constraints (\ref{dq23}) is defined according to the following
equality:
\begin{gather}
[\Phi,\,{\cal F}]^*\equiv[\Phi',\,{\cal F}']\,.
\label{dq24}
\end{gather}
The remarkable property of the considered theory is the fact that
\begin{gather}
[\Phi,\,{\cal H}_T]^*\approx [\Phi,\,{\cal H}_T]\,.
\label{dq25}
\end{gather}
Here the approximate equality means that after the imposition of
all first and second class constraints the operators in the both
sides of Eq. (\ref{dq25}) coincide, that is the weak equality (\ref{dq25}) reduces
to the strong one. Relation (\ref{dq25}) follows immediately from Eqs.
(\ref{dq15}) and (\ref{dq24}). Eq. (\ref{dq25}) means that the Heisenberg equation
$$
i\dot{\Phi}=[\Phi,\,{\cal H}_T]^*
$$
for any field in reduced theory coincides weakly with
corresponding Heisenberg equation in nonreduced theory. Evidently,
this remarkable conclusion retains true under imposition of any
number of pairs of the second class constraints of type
(\ref{dq23}) \footnote{In Appendix B we give the simple example in which the
imposition of second class constraints of type (\ref{dq23}) does not
change equations of motion.}.

The above-stated bring to the following idea of ultraviolet
regularization of quantum theory of gravity. Let a local field
$\Phi(x)$ create and annihilate particles in the states with wave
functions $\{\phi_N(x)\}$ by creation and annihilation operators
$\{A^{\dag}_N,\,A_N\}$ (for simplicity the field $\Phi$ is assumed
to be real). The physical space of states is invariant relative to
the action of creation and annihilation operators. Therefore there
is the possibility of imposing the second class constraints of the
type (\ref{dq23}) for any number of pairs of these operators without
changing Heisenberg equations of motion. Let the high-frequency
(in some sense) wave functions $\{\phi_N(x)\}_{|N|>N_0}$ have the
value of index $|N|>N_0$. The ultraviolet regularization of the
theory is performed by imposing the constraints of the type (\ref{dq23})
for all $|N|>N_0$. It is very important that under the constraints
the regularized equations of motion and first class constraints
preserve their canonical form. Hence the equations of regularized
theory are general covariant, i.e. they conserve their form under
the general coordinate transformations and local frame
transformations.

Since unregularized theory of quantum gravity is mathematically
meaningless, so it seems correct the direct definition of
regularized theory by means of introduction of natural axioms. \\

{\bf {Axiom 1.}} {\it {All states of the theory which are
physically meaningful are obtained from the ground state}} \
$\vert \, 0 \,\rangle $ \ {\it {using the creation operators}} \
$A^{\dag}_N$ \ with \ $\vert \, N \,\vert <N_0$ \ :
 \begin{gather}
 \vert \,n_1,\,N_1;\ldots ;\,n_s,\,N_s\,\rangle=
 \nonumber \\
 =(n_1!\cdot\ldots\cdot
 n_s!\,)^{-\frac{1}{2}} \cdot
 (A^{\dag}_{N_1})^{n_1}\cdot\ldots\cdot
 (A^{\dag}_{N_s})^{n_s}\,\vert \,0\,\rangle \ ,
 \nonumber \\
   A_N\,\vert\,0\,\rangle=0\,.
 \label{dq26}
\end{gather}
{\it{States (\ref{dq26}) form an orthonormal basis of the space \
$F$ \ of physical states of the theory.}} \\

{\bf{Axiom 2.}} {\it{The dynamical variables}} \ ${\Phi}(x)$ \
{\it {transfer state (\ref{dq26}) with fixed values of numbers $(n_1, \,
N_1; \ldots; \, n_s, \, N_s)$ into a superposition of the states
of the theory of form (\ref{dq26}), containing all states in which one of
the occupation numbers is different in modulus by one and all
other
occupation numbers are identical to those of state (\ref{dq26}).}} \\

{\bf{Axiom 3.}} {\it{The equations of motion and constraints for
the physical fields \ $\{\Phi (x), \, {\cal P} (x) \}$ have the
same form, to within the arrangement of the operators, as the
corresponding classical equations and constraints.}} \\

Further we suppose that the momentum variables ${\cal P} (x)$ are
expressed through the fundamental field variables $\Phi (x)$ and
their time derivatives $\dot{\Phi}(x)$, so that the Lagrange
equations instead of Hamilton equations are used.

Let's assume, further, that the ground state $\vert \, 0 \,\rangle$
is a coherent state with respect to the gauge degrees of freedom.
It means that the quantum fluctuations of the gauge degrees of
freedom are not significant and their dynamics in fact is
classical.

Let's emphasize that this assumption is related with the fact of
noncompactness of the gauge group. (Since the group of general
linear transformations is noncompact, so the gauge group in the
theory of gravity is noncompact.)
The quasiclassical charecter of dynamics of gauge degrees of
freedom seems true only for noncompact gauge groups. On the
contrary, the motion in compact gauge group (such as in Yang-Mills
theory) can not be regarded as classical.

Let's consider, for example, the quantized electrodynamic field in
noncovariant Coulomb gauge. In this gauge only the degrees of
freedom describing photons fluctuate, but the gauge (longitudinal)
degrees of freedom are defined unambiguously through the electric
current. Thus, the gauge degrees of freedom in QED does not
fluctuate, effectively they are classical. On the other hand, in
high-temperature confinement phase in QED on a lattice the
high-temperature expansion is valid. In this case the gauge
degrees of freedom can not be regarded as classical. So our
assumption about classical behavior of gauge degrees of freedom in
quantum gravity is equivalent to the assumption that quantum
gravity is in noncompact phase.

Consider any fundamental field:
\begin{gather}
\Phi (x)=\Phi_{(cl)} (x)+ \sum_{|N|<N_0} \,[\,\phi_N
(x)\,A_N+\phi^*_N (x)\,A^{\dag}_N\,]+\ldots \,.
\label{dq27}
\end{gather}
{\it {On the right-hand side of Eq.}} (\ref{dq27}) {\it {all
functions \ $\Phi_{(cl)} (x),\,\, \phi_N (x)$, \ and so on are
$c$-number functions}}.

This follows from the assumption about the quasiclassical
character of the dynamics of gauge degrees of freedom.

Now we can supplement our system of axioms by the following
supposition: field (\ref{dq27}) is used in axioms 1-3.
 The fields $\Phi_{(cl)} (x),\,\phi_N (x),\,\psi_N (x)$,
and so on satisfy certain equations which can be obtained uniquely
from the Lagrange equations of motion, if the expansion of the
field $\Phi (x)$ in form (\ref{dq27}) is substituted into them and
then, after normal ordering of the operators
$\{\,A_N,\,A^{\dag}_N\,\}$, the coefficients of the various powers
of the generators of the Heisenberg algebra
$\{\,A_N,\,A^{\dag}_N\,\}$ are equated to zero. As a result of the
indicated normal ordering, the relations arise between the higher
order coefficient functions and the lower order coefficient
functions in expansion (\ref{dq27}). We obtain an infinite chain
of equations for the coefficient functions $\{\,\Phi^{(cl)}
(x),\,\phi_N (x), \,\psi_N (x),\,\ldots\}$. The latter conjecture
can be
introduced with the aid of the following axiom, replacing axiom 3. \\

{\bf {Axiom $3^{\prime}$.}} {\it {The equations of motion for the
quantized fields}} (\ref{dq27}), {\it {up to the ordering of the
quantized fields, have the same form as the corresponding
classical equations
of motion.}} \\

\section{Dynamic Quantization
of Gravity }

We shall now apply the quantization scheme developed above to the
theory of gravity.

Let's consider the theory of gravity with a $ \Lambda $ term which
is coupled minimally with the Dirac field. The action of such a
theory has the form
\begin{gather}
S=-\frac{1}{l^2_P}\int d^4\,x\,\sqrt{-g}\,\,(R+2\Lambda)+
\nonumber \\
+\int d^4\,x\,\sqrt{-g}\,\,\biggl\{\frac{i}{2}\,e^{\mu}_{a} \Bigl(
\overline{\psi}\gamma^a\,D_{\mu}\psi-\overline{D_{\mu}\psi}\,
\gamma^a\,\psi\Bigr)-m\overline{\psi}\psi\biggr\}\,.
\label{dqg28}
\end{gather}
Here $\{e^{\mu}_a\}$ is an orthonormalized basis, \ $g_{\mu\nu}$ \
is the metric tensor, and \ $\eta_{ab}=diag(1,\,-1,\,-1,\,-1)$ \,
so that
$$
g_{\mu\nu}\,e^{\mu}_a\,e^{\nu}_b=\eta_{ab}, \ \ \ \
R=e^{\mu}_a\,e^{\nu}_b\,R^{ab}_{\mu\nu}\,,
$$
the 2-form of the curvature is given by
$$
d\omega^{ab}+\omega^a_c\wedge\omega^{cb}=
 \frac{1}{2}\,R^{ab}_{\mu\nu}\,dx^{\mu}\wedge dx^{\nu}\,,
$$
where the 1-form \ $\omega^a_b =\omega^a_{b\mu}\,dx^{\mu}$ \ is the
connection in the orthonormal basis \ $\{e^{\mu}_a\}$. The spinor
covariant derivative is given by the formula
$$
D_{\mu}\psi=\left(\frac{\partial}{\partial x^{\mu}}+
\frac{1}{2}\,\omega_{ab\mu}\,\sigma^{ab}\right)\,\psi\,,  \ \
\sigma^{ab}=\frac{1}{4}\,[\gamma^a,\,\gamma^b]\,,
$$
$\gamma^a$  are the Dirac matrices:
$$
\gamma^a\,\gamma^b+\gamma^b\,\gamma^a=2\,\eta^{ab}\,.
$$

Let's write out the equations of motion for system (\ref{dqg28}).

The variation of action (\ref{dqg28}) relative to the connection
gives the equation
\begin{gather}
\nabla_{\mu}\,e^a_{\nu}-\nabla_{\nu}\,e^a_{\mu}=-\frac{1}{4}l^2_P\,
\varepsilon^a_{\,\,bcd}\,e^b_{\mu}e^c_{\nu}\,\overline{\psi}\gamma^5
\gamma^d\,\psi\equiv T^a_{\mu\nu}\,.
\label{dqg29}
\end{gather}
In deriving the last equation, we employed the equality
\begin{gather}
\gamma^a\,\sigma^{bc}+\sigma^{bc}\,\gamma^a=
-i\varepsilon^{abcd}\,\gamma^5\,\gamma_d\,. \label{dqg30}
\end{gather}
Here \ $\varepsilon_{abcd}$ \ is the absolutely antisymmetric
tensor, and \ $ \varepsilon_{0123}=1$. The right-hand side of Eq.
(\ref{dqg29}) is the torsion tensor.

We note that torsion (\ref{dqg29}) possesses the property
\begin{gather}
T^{\nu}_{\mu\nu}\equiv e^{\nu}_a\,T^a_{\mu\nu}\equiv 0\,.
\label{dqg31}
\end{gather}
Consequently, even though torsion exists in the considered theory,
the torsion tensor is not present in the Dirac equation:
\begin{gather}
\bigl(i\,e^{\mu}_a\,\gamma^a\,D_{\mu}-m\bigr)\,\psi=0\,.
\label{dqg32}
\end{gather}

The variation of action (\ref{dqg28}) relative to the orthonormal
basis gives the Einstein equation, which we write in the form
\begin{gather}
R_{\mu\nu}+\Lambda\,g_{\mu\nu}=\frac{1}{2}l^2_P\,\biggl\{
\frac{i}{2}\Bigl(\overline{\psi}\,\gamma^c\,e_{c(\mu}
D_{\nu)}\psi-
\nonumber \\
-e_{c(\mu}\overline{D_{\nu )}\psi}\,\gamma^c\,\psi\Bigr)-
\frac{1}{2}m\,\overline{\psi}\psi\,g_{\mu\nu}\biggr\}\,.
\label{dqg33}
\end{gather}
Here the expression in braces is
$(T_{\mu\nu}-1/2\,g_{\mu\nu}\,T)$, where $T_{\mu\nu}$ is the
energy-momentum tensor on the mass shell (i.e., taking account of
the equations of motion of matter --- in our case, the Dirac
equation (\ref{dqg31})).

Equations (\ref{dqg29}), (\ref{dqg30})--(\ref{dqg33}), together
with the relations
$$
g_{\mu\nu}=\eta_{ab}e^a_{\mu}e^b_{\nu}\,,
\ \ \ e^{\mu}_a\,e^b_{\mu}=\delta^b_a
$$
form a complete system of classical equations of motion and
constraints for system (\ref{dqg28}).

We now represent the fields as the sum of classical and quantum
components:
\begin{gather}
g_{\mu\nu}=g_{(cl)\mu\nu}+h_{\mu\nu}\,\,\,, \,\,\, e^a_{\mu}=
e^a_{(cl)\mu}+f^a_{\mu}\,.
\label{dqg34}
\end{gather}
Assume that the fermion field has no classical component, so that
\begin{gather}
\psi (x)=\sum_{|N|<N_F} \Bigl(a_N\,\psi_N^{(+)}(x)+b_N^{\dag}
\,\psi_N^{(-)}(x)\Bigr)+\ldots \,\,\,,
\label{dqg35}
\end{gather}
where the Fermi creation and annihilation operators satisfy the
following anticommutation relations (as usual, only the nonzero
relations are written out):
\begin{gather}
\{\,a_M\,,\,a_N^{\dag}\,\}=\{\,b_M\,,\,b_N^{\dag}\,\}=
\delta_{M,N}\,,
\nonumber \\
 a_N|0\rangle=b_N|0\rangle=0\,.
\label{dqg36}
\end{gather}
The complete orthonormal set of fermion modes
$\Bigl\{\psi_N^{(\pm)}(x)\Bigr\}$ can be naturally determined as
follows. Denote by $\Sigma^{(3)}$ the spacelike hypersurface,
defined by the equation $t=\Const$, and by $\Sigma^{(3)}_0$ the
hypersurface at $t=t_0$. Let the metric in space-time be given by
means of the tensor $g_{\mu\nu}$. This metric induces a metric on
$\Sigma^{(3)}_0$, which in the local coordinates
$x^i\,,\,i=1,2,3,$ is represented by the metric tensor ${}^3g
_{ij}$. Using the equations
$$
{ }^3g_{ij,k}=\gamma^l_{ik}\,{ }^3g_{lj}+\gamma^l_{jk}\,{ }^3g_{il}\,,
 \ \ \ \gamma^k_{ij}=\gamma^k_{ji}\,,
$$
$$
{ }^3g_{ij}=-\sum^3_{\alpha=1}\,{ }^3e^{\alpha}_i\,{ }^3e^{\alpha}_j\,,
\ \ \ { }^3e^{\alpha}_i\,{ }^3e^i_{\beta}=\delta_{\alpha\beta}\,,
$$
$$
\partial_i{ }^3e^i_{\alpha}+\gamma^j_{ki}\,{ }^3e^k_{\alpha}+
{ }^3\omega_{\alpha\beta i}\,{ }^3e^j_{\beta}=0\,, \ \ \
{ }^3\omega_{\alpha\beta i}=-{ }^3\omega_{\beta\alpha i}
$$
the connection (without torsion) \ $\gamma^i_{jk}$ \ in local
coordinates and a spin connection \ ${ }^3\omega_{\alpha\beta i}$
are determined on \ $\Sigma^{(3)}_0$. For a Dirac single-particle
Hamiltonian we have:
$$
{\cal H}_D=-i\,{ }^3e_{\alpha}^i\,\alpha^{\alpha}\,
(\partial_i+\frac{1}{2}\,{ }^3\omega_{\beta\gamma
i}\,\frac{1}{4}\,
[\alpha^{\beta},\,\alpha^{\gamma}\,]\,)+m\,\gamma^0\,,
$$
$$
\alpha^{\beta}=\gamma^0\,\gamma^{\beta}
$$
It is easy to check that in the metric
\begin{gather}
\langle\,\psi_M,\,\psi_N\,\rangle=\int_{\Sigma^{(3)}_0}\,d^3x\,
\sqrt{-{ }^3g}\,\psi^{\dag}_M\,\psi_N
\label{dqg37}
\end{gather}
the operator \ ${\cal H}_D$ \ is self-conjugated. Consequently,
the solution of the problem for the eigenvalues on \
$\Sigma^{(3)}_0$
\begin{gather}
{\cal
H}^{(0)}_D\,\psi^{(\pm)}_N(x)=\pm\varepsilon_N\,\psi_N^{(\pm)}(x)\,,
\ \ \ \
 \varepsilon_N > 0
\label{dqg38}
\end{gather}
has a complete set of orthonormal modes in metric (\ref{dqg37}). The index
(0) everywhere means that in the corresponding quantity the fields
are taken in the zero approximation with respect to quantum
fluctuations.

Note that a one-to-one relation can be established between the
positive- and negative-frequency modes by means of the equation
$$
\gamma^0\gamma^5\,\psi^{(+)}_M=\psi_M^{(-)}
$$

We call the attention to the fact that the scalar product
\begin{gather}
(\psi_M,\,\psi_N)=\int_{\Sigma^{(3)}}\,d^3x\,\sqrt{-g^{(0)}}\,
\psi^{\dag}_M\,\psi_N
\label{dqg39}
\end{gather}
is not always the same as the scalar product (\ref{dqg37}). These scalar
products coincide, if the path function \ $N=1$, which happens, for
example, for the metric
$$
g^{(0)}_{0i}=0\,, \ \ \ g^{(0)}_{00}=1\,.
$$
The scalar product (\ref{dqg39}) has the advantage over the scalar product
(\ref{dqg37}) that if the modes \ $\{\psi^{(\pm)}_N(x)\,\}$ \ satisfy the
Dirac equation in the zero approximation with respect to quantum
fluctuations (which, according to the exposition below, does indeed
happen), then the scalar product (\ref{dqg39}) is conserved in time.

The field $h_{\mu\nu}$ in Eq. (\ref{dqg34}) can be expanded as follows:
\begin{gather}
h_{\mu\nu}= l_P \sum_{|N|<N_0}(h_{N\,\,\mu \nu} c_N+ h^*_{N\,\,\mu
\nu} c^{\dag}_N)+
\nonumber \\
+l^2_P \Bigl\{\sum_{|N_1|,|N_2|<N_0} (h_{N_1 N_2\,\,\mu
\nu}c_{N_1} c_{N_2}+ h^*_{N_1 N_2\,\,\mu \nu}c^{\dag}_{N_1}
c^{\dag}_{N_2}+
\nonumber \\
+h_{N_1 \mid N_2\,\,\mu \nu}c^{\dag}_{N_1}
c_{N_2})+
\nonumber \\
+\sum_{|N_1|,\,|N_2|<N_F}\,(\,h^{F(++)}_{N_1\,N_2\,\mu\nu}\,
a^{\dag}_{N_1}\,a_{N_2}+
h^{F(--)}_{N_2\,N_1\,\mu\nu}\,b^{\dag}_{N_1}\,b_{N_2}+
\nonumber \\
+h^{F(+-)}_{N_1\,N_2\,\mu\nu}\,a^{\dag}_{N_1}\,b^{\dag}_{N_2}+
h^{F(+-)*}_{N_1\,N_2\,\mu\nu}\,b_{N_2}\,a_{N_1})\Bigr\}+\ldots
\label{dqg40}
\end{gather}
In Eqs. (\ref{dqg34}), (\ref{dqg35}), and (\ref{dqg40}) the $c$-number coefficient fields \
$\psi^{(\pm)}_N\,,\,g_{(cl)\,\mu\nu}$ , \ $h_{N\,\mu\nu}$ \ and so
on can be expanded in powers of the Planck scale, for example
$$
g_{(cl)\,\mu\nu}=g^{(0)}_{\mu\nu}+l^2_p\,g^{(2)}_{(cl)\,\mu\nu}+\ldots
$$
Since fields (\ref{dqg40}) are real, we have
\begin{gather}
h_{N_1 N_2\,\,\mu \nu}=h_{N_2 N_1\,\,\mu \nu}\,,\,\,
h^*_{N_1 \mid N_2\,\,\mu \nu}=h_{N_2 \mid N_1\,\,\mu \nu}\,,
\nonumber \\
h^{F(++)*}_{N_2\,N_1\,\mu\nu}=h^{F(++)}_{N_1\,N_2\,\mu\nu}\,,\,\,
h^{F(--)*}_{N_2\,N_1\,\mu\nu}=h^{F(--)}_{N_1\,N_2\,\mu\nu}
\label{dqg41}
\end{gather}
The operators \ $\{c_N,\,c_N^+\,\}$ \ satisfy the Bose commutation
relations (\ref{dqg31}). A method for choosing the set of functions \
$\{h_{N\,\mu\nu}\,\}$ \ will be discussed below.

According to the dynamic quantization scheme, we must substitute
fields (\ref{dqg34})--(\ref{dqg35}) and (\ref{dqg40}) into Eqs. (\ref{dqg29}) and
(\ref{dqg32})--(\ref{dqg33}), after which the
operators $\{\,A_N,\,A^{\dag}_N\,\}$ must be normal-ordered and all
coefficients of the various powers of these operators and the
Planck scale must be set equal to zero.

Thus, we obtain the first of these equations:
\begin{gather}
\nabla^{(0)}_{\mu}\,e^{(0)\,a}_{\nu}-\nabla^{(0)}_{\nu}\,e^{(0)\,a}_{\mu}=0\,,
\,\,\,
R^{(0)}_{\mu \nu}+\Lambda\,g^{(0)}_{\mu \nu}=0
\label{dqg42}
\end{gather}
Here and below all raising and lowering of indices are done with
the tensors $g^{(0)}_{\mu \nu}$ and $g^{(0)\mu\nu}$. Thus, in the
lowest approximation the fields satisfy the classical equations of
motion. In the zeroth approximation we also have a series of
equations for the fermion modes:
\begin{gather}
\bigl(i\,e^{(0)\,\mu}_a\,\gamma^a\,D^{(0)}_{\mu}-m\,\bigr)\,
\psi^{(0)(\pm)}_N=0
\label{dqg43}
\end{gather}

We now introduce the notation
\begin{gather}
K^{(0) \lambda \rho}_{\mu \nu}=\left [-\frac{1}{2}\nabla^{(0)}_{\sigma}
\nabla^{(0)\sigma}\,\delta^{\lambda}_{\mu}\,\delta^{\rho}_{\nu}-
R^{(0)\lambda\,\,\rho}_{\,\,\,\mu\,\,\,\nu}+
R^{(0)\rho}_{\nu}\,\delta^{\lambda}_{\mu}+\right.
\nonumber \\
\left.+\nabla^{(0)}_{\mu}\,\left (\,\nabla^{(0)\lambda}\,
\delta^{\rho}_{\nu}-\frac{1}{2}\,\nabla^{(0)}_{\nu}\,g^{(0)\lambda \rho}
\right )\,\right ]+
\nonumber \\
+[\,\mu \longleftrightarrow \nu ]+
2\,\Lambda\,\delta^{\lambda}_{(\mu}\delta^{\rho}_{\nu)}\,,
\label{dqg44}
\end{gather}
\begin{gather}
R^{(0)(2)}_{\mu \nu}(h,\,h^{\prime})=\frac{1}{2}\,\bigg[\,R^{(0)(2)}_{\mu \nu}
(h+h^{\prime},\,h+h^{\prime})-
\nonumber \\
-R^{(0)(2)}_{\mu \nu}(h,\,h)-
R^{(0)(2)}_{\mu \nu}(h^{\prime},\,h^{\prime}) \bigg]
\label{dqg45}
\end{gather}
It is easily checked that
$$
\frac{1}{2}\,K^{(0)\,\lambda\rho}_{\mu\nu}=\frac{\delta\,
(R_{\mu\nu}+\Lambda\,g_{\mu\nu})}{\delta\,g_{\lambda\rho}}\,\bigg|_{g_{\mu\nu}=
g^{(0)}_{\mu\nu}}\,,
$$
where $R^{(0)(2)}_{\mu\nu}(h,\,h)$ is a quadratic form of the
tensor field $h_{\lambda\rho}$ which can be constructed in terms of
the second variation of \ $R_{\mu\nu}$ relative to the metric
tensor at the point \ $g^{(0)}_{\mu\nu}$. Let's write out the
complete form:
$$
R^{(0)(2)}_{\mu\nu}(h,h)=
\frac{1}{2}\,(h^{\rho}_{\lambda}\,
h^{\lambda}_{\rho ;\mu}\,)_{;\nu}-
$$
$$
-\frac{1}{2}\,[\,h^{\lambda}_{\sigma}\,(
h^{\sigma}_{\mu;\nu}+h^{\sigma}_{\nu;\mu}-
h^{;\sigma}_{\mu\nu}\,)\,]_{;\lambda}+
$$
$$
+\frac{1}{4}\,h^{\lambda}_{\lambda;\rho}\,
(h^{\rho}_{\mu;\nu}+h^{\rho}_{\nu;\mu}-
h^{;\rho}_{\mu\nu}\,)-
$$
$$
-\frac{1}{4}\,(h^{\lambda}_{\rho;\nu}+
h^{\lambda}_{\nu;\rho}-h^{;\lambda}_{\nu\rho}\,)
\,(h^{\rho}_{\mu;\lambda}+
h^{\rho}_{\lambda;\mu}-h^{;\rho}_{\mu\lambda}\,)
$$
Thus, $R^{(0)(2)}_{\mu\nu}(h,\,h^{\prime})$ \ is a symmetric
bilinear form with respect to its arguments \ $h_{\mu\nu}$ \ and \
$h^{\prime}_{\lambda\rho}$, which in what follows are operator
fields (\ref{dqg40}). Thus, here the problem of ordering the operator fields
to lowest order has been solved.

Now we can write out the following relations, which follow from the
exact quantum equations with the expansion indicated above. To
first order in \ $l_P$ \ we have
\begin{gather}
\frac{1}{2}\,K^{(0)\,\lambda \rho}_{\mu \nu}\,h_{N\,\lambda \rho}=0\,.
\label{dqg46}
\end{gather}

We note that, using Eqs. (\ref{dqg42}), the operator (\ref{dqg44}) vanishes on the
quantity \ $(\xi_{\mu\,;\nu}+\xi_{\nu\,;\mu})$. Consequently, the
value of the operator (\ref{dqg44}) on the fields \ $h_{\mu\nu}$ \ and
\begin{gather}
h^{\prime}_{\mu\nu}=h_{\mu\nu}+\xi_{\mu\,;\nu}+\xi_{\nu\,;\mu}
\label{dqg47}
\end{gather}
coincide for any vector field \ $\xi_{\mu}$. This fact is a
consequence of the gauge invariance of the theory. Using the
indicated gauge invariance, any solution of Eq. (\ref{dqg46}) can be put
into the form
\begin{gather}
\nabla^{(0)}_{\nu}\,h^{\nu}_{\mu}-\frac{1}{2}\,
\nabla^{(0)}_{\mu}\,h^{\nu}_{\nu}=0\,.
\label{dqg48}
\end{gather}
In what follows, we shall assume that the field satisfies the gauge
condition (\ref{dqg48}), which is convenient in a number of problems. It is
obvious that taking account of the gauge condition (\ref{dqg48}) the terms
in round brackets in operator (\ref{dqg44}) vanishes.

To clarify the question of the normalization of the gravitational
modes, we shall employ the following technique. The equation of
motion (\ref{dqg46}) can be obtained with the help of the action
\begin{gather}
S^{(2)}=\int\,d^4x\,\sqrt{-g^{(0)}}\,
h^{\mu\nu}\,K^{(0)\,\lambda\rho}_{\mu\nu}\,h_{\lambda\rho}\,.
\label{dqg49}
\end{gather}
Hence follows the canonically-conjugate momentum for the field
\ $h_{\mu\nu}$ \ and the simultaneous commutation  relations:
\begin{gather}
\pi^{\mu\nu}=\sqrt{-g^{(0)}}\,\nabla^{(0) 0}\,h^{\mu\nu}\,,
\nonumber \\
[h_{\mu\nu}(x),\,\pi^{\lambda\rho}(y)\,]=
i\,\delta^{\lambda}_{(\mu}\,\delta^{\rho}_{\nu )}\,\delta^{(3)}(x-y)\,.
\label{dqg50}
\end{gather}
Evidently, in Eq. (\ref{dqg50}) the fields are free of constraints (\ref{dqg48}).
Let's represent the field \ $h_{\mu\nu}$ \ in the form (compare
with the first term in Eq. (\ref{dqg40}))
\begin{gather}
h_{\mu\nu}(x)=\sum_N\,\bigl(h_{N\,\mu\nu}(x)\,c_N+
h^*_{N\,\mu\nu}(x)\,c^{\dag}_N\bigr)\,.
\label{dqg51}
\end{gather}
The set of operators \ $\{c_N,\,c^{\dag}_N \}$ \ form a Heisenberg
algebra, and the functions \ $\{h_{N\,\mu\nu}\}$ \ satisfy Eq.
(\ref{dqg46}). Equations (\ref{dqg50}) and (\ref{dqg51}) lead to the following relations
reflecting the orthonormal nature of the set of the modes:
\begin{gather}
i\,\int_{\Sigma^{(3)}}\,d^3x\,\sqrt{-g^{(0)}}\,
\bigl[h^{\mu\nu *}_M\,\nabla^{(0)\,0}\,h_{N\,\mu\nu}-
\nonumber \\
-(\nabla^{(0)\,0}\,h_M^{\mu\nu\,*})\,h_{N\,\mu\nu}\,\bigr]=
\delta_{M,\,N}\,.
\label{dqg52}
\end{gather}
In the latter equations the integration extends over any spacelike
hypersurface \ $\Sigma^{(3)}$. As a result of Eqs. (\ref{dqg46}), integrals
(\ref{dqg52}) indeed do not depend on the hypersurface. It is natural to
assume that the gravitational modes satisfy conditions (\ref{dqg52}). The
significance of Eqs. (\ref{dqg52}) is that the normalization of the
coefficient functions in expansion (\ref{dqg40}) is given with its help.

In the second order in $l_P$, we obtain the following equations:
\begin{gather}
\frac{1}{2}\,K^{(0)\,\lambda \rho}_{\mu \nu}\,h_{N_1 N_2\,\lambda \rho}
=-R^{(0)(2)}_{\mu \nu}\,(h_{N_1},\,h_{N_2})\,\,,
\label{dqg53}
\end{gather}
\begin{gather}
\frac{1}{2}\,K^{(0)\,\lambda \rho}_{\mu \nu}\,
h_{N_1 \mid N_2\,\lambda \rho}=
-2\,R^{(0)(2)}_{\mu \nu}\,(h_{N_1}^*,\,h_{N_2})\,\,,
\label{dqg54}
\end{gather}
\begin{gather}
\frac{1}{2}\,K^{(0)\,\lambda \rho}_{\mu \nu}\,
h^{F(\pm\pm)}_{N_1\,N_2\,\lambda \rho}=\pm\frac{i}{4}\,
\Bigl(\overline{\psi}^{(0)(\pm)}_{N_1}\,\gamma^c\,e^{(0)}_{c(\mu}\,
D^{(0)}_{\nu)}\,\psi^{(0)(\pm)}_{N_2}-
\nonumber \\
-e^{(0)}_{c(\mu}\overline{D_{\nu )}^{(0)}\,
\psi^{(0)(\pm)}_{N_1}}\,\gamma^c\,\psi^{(0)(\pm)}_{N_2}\,\Bigr)\,,
\label{dqg55}
\end{gather}
\begin{gather}
\frac{1}{2}\,K^{(0)\,\lambda \rho}_{\mu \nu}\,
h^{F(+-)}_{N_1\,N_2\,\lambda \rho}=\frac{i}{4}\,
\Bigl(\overline{\psi}^{(0)(+)}_{N_1}\,\gamma^c\,e^{(0)}_{c(\mu}\,
D^{(0)}_{\nu)}\,\psi^{(0)(-)}_{N_2}-
\nonumber \\
-e^{(0)}_{c(\mu}\overline{D_{\nu )}^{(0)}\,
\psi^{(0)(+)}_{N_1}}\,\gamma^c\,\psi^{(0)(-)}_{N_2}\,\Bigr)\,,
\label{dqg56}
\end{gather}
\begin{gather}
\frac{1}{2}\,K^{(0)\,\lambda \rho}_{\mu \nu}\,g^{(2)}_{(cl)\,\lambda \rho}=
-\sum_{|N|<N_0}\,R^{(0)(2)}_{\mu \nu}\,(h_N^*,\,h_N)+
\nonumber \\
+\frac{i}{4}\,\sum_{|N|<N_F}\,\Bigl(\overline{\psi}^{(0)(-)}_{N}\,\gamma^c\,
 e^{(0)}_{c(\mu}\,
D^{(0)}_{\nu)}\,\psi^{(0)(-)}_{N}-
\nonumber \\
-e^{(0)}_{c(\mu}\overline{{\cal
D}_{\nu )}^{(0)}\,
\psi^{(0)(-)}_{N}}\,\gamma^c\,\psi^{(0)(-)}_{N}\,\Bigr)\,.
\label{dqg57}
\end{gather}
It is evident from Eq. (\ref{dqg29}) that torsion appears in the same order
\ $\sim l^2_P$. Here, however, we do not write out the
corresponding corrections for the connection.

We shall now briefly summarize the results obtained.

According to the dynamic quantization method, the quantization of
gravity starts with finding a solution of the classical
microscopic field equations of motion (for example, the solution
of Eqs. (\ref{dqg42}) in the example considered above). The classical
solution is determined by (or determines) the topology of
space-time. Then, using the classical approach, Eqs. (\ref{dqg43}) and
(\ref{dqg46}), which determine the single-particle modes
$\{\,\psi^{(\pm)}_N\,,\,h_{N\,\mu \nu}\}$, are solved. To solve
Eq. (\ref{dqg46}) the gauge must be fixed, since the operator (\ref{dqg44}) is
degenerate because of the gauge invariance of the theory. At the
first step these modes are determined in the zeroth approximation
according to the Planck scale, and their normalization is fixed
using Eqs. (\ref{dqg39}) and (\ref{dqg52}). Given the set of modes
$\{\,\psi^{(0)(\pm)}_N\,,\,h_{N\,\mu \nu}\}$, we can explicitly
write out the right-hand sides of Eqs. (\ref{dqg53})--(\ref{dqg57}) and then solve
them for the two-particle modes $h_{N_1\,N_2\,\,\mu \nu}$,
$h_{N_1\,\mid N_2\,\,\mu \nu}$, and so on, and find the correction
$g^{(2)}_{(cl)\,\mu \nu}$ which is of second order in $l_P$ to the
classical component of the metric tensor. We call attention to the
fact that the right-hand side of Eq. (\ref{dqg57}) arises because the
operators must be normal-ordered. The solution of Eq. (\ref{dqg57}) can be
interpreted as a single-loop contribution to the average of the
metric tensor with respect to the ground state.

We note that if a nonsymmetric bilinear form were used on the
right-hand sides of Eqs. (\ref{dqg53})--(\ref{dqg57}), then the condition that the
metric tensor be real would be violated. Consequently, the
condition that the metric tensor is real determines the ordering of
the operator fields in the equations of motion at least in second
order with respect to the operator fields.

It is important that all Eqs. (\ref{dqg42}), (\ref{dqg46}), and so on which arise are
generally covariant, since they are expansions of generally
covariant equations. Thus, the dynamic quantization method leads to
a regularized gauge-invariant theory of gravity, which contains an
arbitrary number of physical degrees of freedom.

We shall now make a remark about the compatibility of Eqs.
(\ref{dqg53})--(\ref{dqg57}) and the analogous equations arising in higher orders. Let
$h_{\mu\nu}$ be an arbitrary symmetric tensor field and $K^{(0)}$
the operator (\ref{dqg44}), acting on this tensor field. It is easily
verified that, using Eqs. (\ref{dqg42}), we obtain the identity (compare
with Eq. (\ref{dqg48}))
\begin{gather}
\nabla^{(0)}_{\nu}\,(K^{(0)}h)^{\nu}_{\mu}-
\frac{1}{2}\nabla^{(0)}_{\mu}\,(K^{(0)}h)^{\nu}_{\nu}=0\,.
\label{dqg58}
\end{gather}
Consequently, in order for Eqs. (\ref{dqg53})--(\ref{dqg57}) to be compatible the
right-hand sides of these equations must satisfy the same
identity. It is easy to see that this is indeed the case in lowest
order. Indeed, Eqs. (\ref{dqg53})--(\ref{dqg56}) are identical to the analogous
classical equations arising when nonuniform modes (higher order
harmonics) and the subsequent expansion of the classical Einstein
equation in powers of the nonlinearity or the Planck length are
added to the uniform fields. Hence it follows that each term on
the right-hand sides of the "loop" equations of the type (\ref{dqg57})
likewise satisfy the necessary identity, since these terms have
the same form as the right-hand sides of the "nonloop" Eqs.
(\ref{dqg53})--(\ref{dqg56}).

In highest orders in creation and annihilation operators the
compatibility of arising equations follows from the gauge
invariance of the regularized Einstein equation. Indeed, the
identity (\ref{dqg58}) is the consequence of gauge invariance (invariance
relative to the general coordinate transformations) of the
equation. To clarify the quation let's rewrite the action (\ref{dqg28})
(for simplicity with $m=0, \, \Lambda=0$) in the following form:
\begin{gather}
S=S_g+S_{\psi}\,,
\label{dqg59}
\end{gather}
\begin{gather}
S_g=-\frac{1}{4l^2_P}\int\d^4
x\varepsilon_{abcd}\varepsilon^{\mu\nu\lambda\rho}e^a_{\mu}R^{bc}_{\nu\lambda}e^d_{\rho}\,,
\label{dqg59 a}
\end{gather}
\begin{gather}
S_{\psi}=\frac16\int\d^4
x\varepsilon_{abcd}\varepsilon^{\mu\nu\lambda\rho}\left[\frac{i}{2}\opsi
e^b_{\nu}e^c_{\lambda}e^d_{\rho}\gamma^aD_{\mu}\psi+h.c.\right]\equiv
\nonumber \\
\equiv\int\d^4 x\opsi\stackrel{\leftrightarrow}{\cal D}\psi\,.
\label{dqg59 b}
\end{gather}
Here $\stackrel{\leftrightarrow}{\cal D}$ is Dirac hermithian
operator, depending on other operator fields. The
Heisenberg--Dirac equations are written in the form
\begin{gather}
\stackrel{\rightarrow}{\cal D}\psi=0\,, \qquad
\opsi\stackrel{\leftarrow}{\cal D}=0\,.
\label{dqg60}
\end{gather}
In Eqs. (\ref{dqg60}) the disposition of creation and annihilation
operators is the same as in the action (\ref{dqg59}). Einstein equation is
the condition of stationarity of the action (\ref{dqg59}) relative to
variations of metric or tetrad. Evidently, the action (\ref{dqg59}) is
invariant under the general coordinate transformation even if the
fields are quantized. This follows from the facts that under the
coordinate transformations all fundamental fields transform
linearly and that the action (\ref{dqg59}) is a polynomial relative to the
fundamental fields. Therefore, if the material fields are on mass
shell (in our case this means that Eqs. (\ref{dqg60}) hold), the action
(\ref{dqg59}) is stationary under infinitesimal gauge transformation of
tetrad field only. This means that the quantum energy-momentum
tensor on mass shell  satisfies to some identity which in
classical limit transforms to the well known identity
$T^{\mu}_{\nu;\,\mu}=0$. From this quantum identity it follows
that if some quantum tetrad field satisfies Einstein equation,
then the field transformed by infinitesimal gauge transformation
also satisfies Einstein equation. From here the compatibility of
quantum Einstein equation follows, as well as the compatibility of
the chain of equations described above. However, this conclusion
is true only if quantum Dirac equations (\ref{dqg60}) hold, and the
operators in the action and energy-momentum tensor are placed so
as in Eq. (\ref{dqg59}). In other words, the creation and annihilation
operators in Eqs. (\ref{dqg59}) and (\ref{dqg60}) must be placed identically. This
is the guarantee of self-consistency of the chain of equations
arising from exact quantum Einstein and motion equations.

We also call attention to the fact that in the dynamic
quantization method it is implicitly assumed that the quantum
anomaly is absent in the algebra of the first class constraints
operators. Consequently, the dynamic quantization method must be
justified in each specific case by concrete calculations, which
must be not only mathematically correct but also physically
meaningful.

\section{The Fundamental Fields and the Secondary Quantized fields}

We see that the fundamental or cosmological fields are expanded in
modes the number of which is finite in the case of compact space.
Packing of the modes is essentially noncompact in momentum space.
Assume that at present at low energies the minimal difference
between the momenta of modes is of the order of $\Delta
k_{min}\sim 1/\lambda_{\max}$. From the consideration at the end
of Section 2 (see Eq. (\ref{discr510})) it follows that
\begin{gather}
\lambda_{\max}\sim\left(\frac{a_0}{a}\right)^{(\ln
3\lambda)/3\lambda}a\ll a\,.
\label{ffsq10}
\end{gather}
Further we denote by $a(t)$ the radius of universe and by $t_0$
the age of universe.

Further, one can assume that in considered theory the
stochastization of phases of modes takes place on distances less
than $\lambda_{\max}$. Under the phase stochastization we mean
that any correlation between phases of wave packets spased by an
enough distance can not take place. Such stochastization must
occur if considered theory is the long-wavelength limit of
discrete quantum theory of gravity discussed in Section 2. The
point is that in long-wavelength limit the lattice action $S$
transforms to the action which is expressed as follows:
$$
S=S_{Einstein}+\Delta S\,.
$$
Here $S_{Einstein}$ is standard Einstein action which does not
retains any information about the structure of lattice, and
$\Delta S$ depends only on higher derivatives of the fields and
also it essentially depends on the structure of the lattice.
Therefore equations of motion contain the items with higher
derivatives of fields and casual coefficients depending on
structure of irregular lattice (simplicial complex). These items
play negligible part for low frequencies modes but their part
increase with increasing of mode frequency. The items with higher
derivatives of fields and casual coefficients lead to diffusional
propagation of modes and so to stochastization of phase on large
distances. But just due to this circumstance the high frequency
wave packets can be localized in relatively small regions of
space. This means that noncompact "packing" of modes in momentum
space does not affects to the possibility of localization of high
frequency wave packets.

One should pay attention to the fact that in presented theory with
noncompact packing of modes the gravitational and gauge
interaction forces does not become weaker. It is seen from quantum
equations of motion which have the canonical form with usual
interaction constants. Thus the interaction between any modes has
the usual strength.

Let's consider, for example, the system of finite number of
electrons, positrons and photons with wavelengths much less than
$\lambda_{max}$. Assume that we are interested in the usual
problem of particle physics: the scattering matrix problem. The
dynamics of real relativistic particles is described by the usual
Dirac and Maxwell equations. The dynamic process of particles
localized in finite space volume $v\ll\lambda^3_{max}$ is studied.
Since the matrix elements between localized states and
nonlocalized states tends to zero as $a^{-3/2}(t_0)$, so only
matrix elements between localized states are significant in the
studied problem. This conclusion is true also with respect to
virtual modes. From here it follows that for description of
processes proceeding in finite volume of space $v$, one must use
the renormalized or secondary quantized quantum fields
$(\psi_r,\,\ldots)$ which are normalized to the volume $v$. This
means that the wave functions of the states
$\{\psi_{r\,N}(x),\ldots\}$ which create and annihilate the
localized particles
 are normalized to
the volume $v$, these wave functions form the complete set of one
particle wave functions with confined energies, and the
corresponding creation and annihilation operators satisfy to
standard relations (\ref{dqg36}). Note that the quantization
conditions (\ref{dqg36}), i.e. nullification of ground state by
annihilation operators, follow from the fact that the causal
correlators $\langle 0\,|\,T\,\psi(x)\,\opsi(y)\,|0\,\rangle$
describe propagation omly positive-frequency waves. It seems that
more general and correct definition of ground state $|0\rangle$
instead of definition (\ref{dq26}) or (\ref{dqg36}) is that the
amplitudes
\begin{gather}
\langle 0\,|\,T\,\psi(x)\,\opsi(y)\,|\,0\,\rangle\,, \qquad
\langle 0\,|\,T\,A_i(x)\,A_j(y)\,|0\,\rangle\,,\,\ldots
\label{ffsq86}
\end{gather}
describe the propagation of only positive-frequency waves if the
times $x^0$ and $y^0$ are close to the time $t_0$. Again the
definition of vacuum depends on the moment of time $t_0$. At
present the state of Universe is close to the ground state. One
can say that the renormalized fields $(\psi_r,\,\ldots)$ are the
secondary quantized fields with the complete (at confined
energies) and normalized on volume $v$ set of one-particle states
$\{\psi_{r\,N}(x),\ldots\}$. Thus the cosmological fields
$(\psi,\,\ldots)$ from which quantum global Einstein equation is
composed and the secondary quantized fields $(\psi_r,\,\ldots)$
are different though they describe the particles with the same
quantum numbers. The causal correlators constructed from
renormalized fields $\psi_r$ (renormalized correlators) and thus
describing local interactions also satisfy the conditions
(\ref{ffsq86}). Since local states normalized to volume $v$ have
compact "packing" in momentum space (at least for experimentally
tested momenta), the renormalized correlators satisfy the standard
equations:
$$
(i\gamma^{\mu}\partial_{\mu}-m)\,\langle
0\,|\,T\,\psi(x)\,\opsi(y)\,|\,0\,\rangle=i\,\delta^4(x-y)\,,
\ldots\,,
$$
which are true at $|x^0-y^0|\gg l_P, \ |\bx-\by|\gg l_P$. And
since the calculations of $S$-matrix elements are performed by
using the standard Dirac and Maxwell equations with usual value of
charge and others parameters, as a result the usual expressions
for $S$-matrix elements are obtained.

Let's emphasize that at solving cosmological problems the retarded
Green functions are used but at calculating $S$-matrix elements
the causal or Feynman one are used.

Does the Casimir effect survives in proposed theory? The answer to
this question is positive. Indeed, the attraction force between
plates of condenser which is caused by Casimir effect is the
derivative of sum of photon zero-point energies with respect to
distance between plates. But only modes with wavelength
commensurable with the distance between plates $d$ really give the
contribution in this derivative. And since $d\ll\lambda_{max}$ the
distortion of Casimir effect does not occurs because this
interaction is described by the secondary quantized fields.

We make the last remark about violation of Lorentz invariance in
the theory. Since as a matter of fact the regularization is
performed here by energy but not Lorentz invariant square of
4-momentum, so Lorentz invariance can be violated. However, until
the processes with low energies (in comparison with the cutoff
energy) are studied the violation of Lorentz invariance is
negligible. The regularization by energy of calculations in QED is
used, for example, in \cite{18}, and at the same time Lorentz
invariance is not violated at low energies. Therefore the fact
that all observed phenomena in nature are Lorentz covariant does
not contradicts to the proposed theory since these phenomena has
been observed at confined energies.

\section{The Possible Solution
of Cosmological Constant Problem}

It follows from Eqs. (\ref{dqg43}) that in lowest order the Dirac
field
\begin{gather}
 \psi^{(1)}(x)=\sum_{|N|<N_0}\left(a_N\psi^{(0)(+)}_N(x)+b^{\dag}_N\psi^{(0)(-)}_N(x)
\right)
\label{cc61}
\end{gather}
satisfies the Dirac equation
\begin{gather}
(ie_a^{(0)\,\mu}\gamma^aD_{\mu}^{(0)}-m)\psi^{(1)}(x)=0\,.
\label{cc62}
\end{gather}
 Here $D_{\mu}^{(0)}$ is the
covariant derivative operator in zeroth order:
\begin{gather}
D_{\mu}^{(0)}=\partial/\partial
x^{\mu}+(1/2)\omega^{(0)}_{ab\mu}\sigma^{ab}+ieA^{(0)}_{\mu}\,,
\label{cc63}
\end{gather}
and $A_{\mu}$ is the gauge field.

It follows from Eq. (\ref{cc62}) that the charge
\begin{gather}
Q=\int\d^3x\sqrt{-g^{(0)}}e^{(0)0}_a(\opsi^{(1)}\gamma^a\psi^{(1)})
\label{cc64}
\end{gather}
conserves (compare with (\ref{dqg39})).

According to (\ref{dqg33}) the contribution of the Dirac field to
the energy-momentum tensor in lowest order is equal to
\begin{gather}
T^{(2)}_{\psi\,\mu\nu}=\Re[i\opsi^{(1)}\gamma^ae^{(0)}_{a(\mu}D_{\nu)}^{(0)}\psi^{(1)}]\,.
\label{cc65}
\end{gather}
Using Eqs. (\ref{dqg36}) it is easy to find vacuum expectation
value of the quantity (\ref{cc65}):
\begin{gather}
\langle
T^{(2)}_{\psi\,\mu\nu}\rangle_0=\Re\left[i\sum_{|N|<N_0}\opsi_N^{(0)(-)}\gamma^ae^{(0)}_{a(\mu}
D^{(0)}_{\nu)}\psi^{(0)(-)}_N\right]\,.
\label{cc66}
\end{gather}

Now let us take into account that the scenario described by the
inflation theory is realized in Universe. It follows from here in
conjunction with the used quantization method that in zeroth
approximation the metric is expressed as
\begin{gather}
\d s^{(0)\,2}=\d t^2-a^2(t)\,\d\Omega^2\,,
\label{cc67}
\end{gather}
where $\d\Omega^2$ is the metric on unite sphere $S^3$, and $a(t)$
is the scale factor of Universe at the running moment of time $t$.
It follows from (\ref{cc67}) that $e_a^{(0)0}=\delta^0_a$ and
$\sqrt{-g^{(0)}}\d^3x=\d V^{(0)}(t)$, where $\d V^{(0)}(t)$ is the
volume element of 3-space in the running moment of time. From
conservation of operator (\ref{cc64}) the conservation of the set
of integrals
\begin{gather}
\int\d
V^{(0)}(t)\,\psi^{(0)(\pm)\dag}_N\psi^{(0)(\pm)}_M=\delta_{NM}
\label{cc68}
\end{gather}
follows. The equality to unity of integrals (\ref{cc68}) means
that the wave functions $\psi^{(0)(\pm)}_M$ are normalized
relative to the volume of all Universe, so that the charge
operator has the form
\begin{gather}
Q=\sum_{|N|<N_0}(a^{\dag}a_N+b_Nb_N^{\dag})\,.
\label{cc69}
\end{gather}

The idea how the vacuum expectation value of the matter
energy-momentum tensor becomes enough small at present is
demonstrated by the following estimation.

According to (\ref{cc68}) we have:
\begin{gather}
\left|\opsi_N^{(0)(\pm)}\psi_N^{(0)(\pm)}\right|\sim\frac{1}{a^3(t)}\,.
\label{cc70}
\end{gather}
Therefore the estimation for the value (\ref{cc66}) is the
following:
\begin{gather}
\langle
T^{(2)}_{\psi\,\mu\nu}\rangle_0\sim\frac{N_0k_{max}}{a^3(t)}\,,
\label{cc71}
\end{gather}
where $k_{max}$ is the value of the order of maximal momentum of
the modes $\{\psi_N^{(0)(\pm)}\}$. It is naturally to suppose that
\begin{gather}
k_{max}\sim l^{-1}_P\sim G^{-1/2}\sim 10^{33}{cm}^{-1}\,.
\label{cc72}
\end{gather}
Since the numerator in the right hand side of relation
(\ref{cc71}) is finite and the denominator is proportional to the
volume of Universe which swells up approximately $10^{100}$ times
more according to inflation scenario, the quantity (\ref{cc71})
can be found enough small at present.

On the other hand, it is seen from the estimation (\ref{cc71})
that at early stages of Universe evolution the quantum
fluctuations played decisive role because the scale of the
Universe were small.

One should pay the attention to the fact that the dynamics of the
system creates two opposite tendencies for mode frequencies
changing.

According to the first tendency the frequencies $\omega$ of all
one-particles modes change in time according to the low
\begin{gather}
\omega\sim\frac{1}{a(t)}\,.
\label{cc73}
\end{gather}
The low (\ref{cc73}) is valid in relativistic case. So, all
frequencies decrees with expansion of Universe.

Now let us write out the first items of formal solution of Dirac
equation (\ref{dqg32}) or (\ref{dqg60}) neglecting gravity degrees
of freedom (i.e. in the case of flat space-time) but in the
presence of gauge field:
\begin{gather}
\psi(x)=\psi^{(1)}(x)+e\int\d^4y\,S_{ret}(x-y)\,A^{(1)}_{\mu}(y)
\gamma^{\mu}\psi^{(1)}(y)+
\nonumber \\
+4\pi\,e^2\int\!\!\int\d^4y\d^4z\,S_{ret}(x-y)\,D_{ret}(y-z)\times
\nonumber \\
\times\left(\opsi^{(1)}(z)
\gamma_{\mu}\psi^{(1)}(z)\right)\gamma^{\mu}\psi^{(1)}(y)+\ldots\,,
\label{cc74}
\end{gather}
\begin{gather}
\left(i\gamma^{\mu}\partial_{\mu}-m\right)S_{ret}(x)=\delta^{(4)}(x)\,,
\label{cc75}
\end{gather}
\begin{gather}
\partial_{\mu}\partial^{\mu}D_{ret}(x)=\delta^{(4)}(x)\,.
\label{cc76}
\end{gather}
Here $S_{ret}(x)$ and $D_{ret}(x)$ are the retarded Green
functions satisfying Eqs. (\ref{cc75}) and (\ref{cc76}). It is
seen from the solution (\ref{cc74}) that the exact field $\psi(x)$
have much more nonzero Fourier components than the field
$\psi^{(1)}$. Hence, the exact solution of quantum Dirac equation
has all Fourier components despite the field of first
approximation $\psi^{(1)}$ has Fourier components only with finite
momenta. From here we see the opposite dynamic tendency: the
frequencies of modes effectively increases as a consequence of
interaction. If the fact of strict conservation of the charge is
taken into account \footnote{It means strict conservation of
quantum charge operator which in lowest order transforms into
expression (\ref{cc64})}, the conclusion about noncompact
"packing" of modes in momentum space should be made. Indeed, let's
calculate the mean value of charge operator relative to a state
$|\,\rangle$. We have:
\begin{gather}
\int\d^3x\langle
\,|\,\psi^{\dag}(x)\psi(x)\,|\;\rangle=\int\frac{\d^3k}{(2\,\pi)^3}
\langle \,|\,\psi^{\dag}_{|\bk|}\psi_{|\bk|}\,|\;\rangle=\const\,,
\nonumber
\end{gather}
\begin{gather}
 \psi_{|\bk|}=\int\d^3x\,e^{-i\bk\bx}\psi(x)\,.
 \nonumber
\end{gather}
 The last relation
means also that the integral
\begin{gather}
\int\d^3x\,\tr\langle
\,|\,T\,\psi(x)\opsi(y)\,|\;\rangle\,\gamma^0\,,
\nonumber
\end{gather}
\begin{gather}
y\longrightarrow x\,, \qquad y^0>x^0
\nonumber
\end{gather}
constructed with the help of correlator $\langle
\,|\,T\,\psi(x)\opsi(y)\,|\;\rangle$ is conserved in time. But the
mean value of energy-momentum tensor is constructed with the help
of the same correlator. From here it is seen the effect of
"loosening of mode packing" in momentum space. This effect is
absent in the theory with dense packing of modes in all diapason
of momenta since all states in momentum space are filled by the
corresponding modes.

To solve the problem of "loosening of mode packing" in momentum
space one must solve quantum kinetic equation for state density in
momentum space. This problem is not solved in this work. But the
fact of noncompact "packing" of modes in momentum space plays an
important role in our consideration. Thus, the noncompact
"packing" of modes in momentum space is taken here as an
assumption.

At a dense "packing" of modes in momentum space the neighbouring
momenta differ by the quantity of the order of $\Delta k_{min}\sim
1/a(t)$. Therefore
\begin{gather}
\d N\sim\frac{a^3(t)\d^3k}{(2\,\pi)^3}\,.
\label{cc77}
\end{gather}
At noncompact "packing" of modes in momentum space the
neighbouring momenta differ by the greater quantity. Assume that
at small momenta this difference at present is of the order of
$\Delta k_{min}\sim 1/\lambda_{\max}$. Furthermore, we shall use
Lorentz-invariant measure in momentum space $[\d^3k/|\bk|]$. Thus
we obtain instead of (\ref{cc77}) the following estimation for the
total number of physical degrees of freedom:
\begin{gather}
N_0\sim\lambda^3_{max}\int^{k_{max}}\frac{\d^3k}{(2\pi)^3(\lambda_{max}|\bk|)}\sim
(\lambda_{max}k_{max})^2\,.
\label{cc78}
\end{gather}
Now using (\ref{cc71}) and (\ref{cc78}) we find:
\begin{gather}
16\pi G\langle
T_{\mu\nu}\rangle_0\sim\frac{l_P^2\,\lambda^2_{max}\,k^3_{max}}{a^3(t)}\leq\Lambda\,,
\label{cc79}
\end{gather}
and from here
\begin{gather}
a(t_0)\geq\frac{(l_P\,\lambda_{max})^{2/3}\,k_{max}}{\Lambda^{1/3}}\,.
\label{cc80}
\end{gather}
If one assume that
\begin{gather}
\lambda_{max}\sim 10^{24}cm\sim 10^{-4}L\,,
\label{cc81}
\end{gather}
where $L=10^{28}cm$ (the dimension of observed part of Universe),
then with the help of relations (\ref{introduction20}),
(\ref{introduction40}), (\ref{cc81}) and (\ref{cc80}) we find the
following estimation for the present dimension of Universe:
\begin{gather}
a(t_0)\geq 10^{17}L\,.
\label{cc82}
\end{gather}

At obtaining the estimation (\ref{cc82}) it was assumed that the
fundamental field theory is not supersymmetric. If one assume that
the fundamental theory is supersymmetric, but the spontaneous
breaking of supersymmetry occurs on the momentum $\sim k_{SS}$,
then the estimation of the dimension of Universe is changed.
Indeed, in this case instead of (\ref{cc72}) we have
\begin{gather}
k_{max}\sim k_{SS}\,,
\label{cc83}
\end{gather}
since according to (\ref{introduction90}) and
(\ref{introduction100}) the boson and fermion contributions to the
vacuum expectation value of energy-momentum tensor with momenta
greater than $k_{SS}$ are mutually cancelled. Therefore instead of
(\ref{cc80}) we obtain:
\begin{gather}
a(t_0)\geq\frac{(l_P\,\lambda_{max})^{2/3}\,k_{SS}}{\Lambda^{1/3}}
\sim10^{29}cm\sim10L\,.
\label{cc84}
\end{gather}
At obtaining the numerical estimation of right hand side Eq.
(\ref{cc84}) we used assumptions (\ref{cc81}) and the popular
assumption in particle physics that $k_{SS}\sim 10^3 GeV\sim
10^{17}cm^{-1}$.

The inclusion of quantum fluctuations of others fields into our
estimations does not changes the result. This is clear already
from dimensional considerations.

The inclusion of higher order corrections by perturbation theory
also does not changes the obtained estimations. Indeed, all known
fundamental interactions except for gravitational are
renormalizable and thus can be considered by perturbation theory
without changing fundamental properties of the vacuum. But the
gravitational quantum corrections are obtained by expanding in
Planck scale $l_P$. Again from dimensional considerations it is
clear that such corrections at passing to the following order in
our theory have the comparative value
\begin{gather}
\sim\left(\frac{l_P}{a(t_0)}\right)^2N_0\sim
\nonumber \\
\sim\left(\frac{l_P}{a(t_0)}\right)^2(\lambda_{max}
k_{max})^2\leq\left(\frac{\lambda_{max}}{a(t_0)}\right)^2\ll 1\,.
\label{cc85}
\end{gather}

\begin{acknowledgments}

I thank participants of seminar of prof. A. A. Belavin for useful
discussion, and especially prof. B. G. Zakharov.  This work was
supported by the Program for Support of Leading Scientific Schools
$\sharp$ 2044.2003.2. and RFBR $\sharp$ 04-02-16970-a.

\end{acknowledgments}

\appendix

\section{}

Let us consider the discrete Laplace operator on a one dimensional
cycle with 3 vertexes (see fig. 1). The numbers $a$, $b$, $c$
are the distances between the vertexes 1 and 2, 2 and 3, 3 and 1,
correspondingly. In the vertexes 1, 2 and 3 the real numbers
$\varphi_1$, $\varphi_2$ and $\varphi_3$ are defined. Write out the discrete
equation for Laplace operator eigenfunctions:
\begin{gather}
-(\Delta\varphi)_1=-\frac{2}{ac}\left(\frac{a\,\varphi_3+c\,\varphi_2}{a+c}
-\varphi_1\right)=\epsilon\,\varphi_1\,,
\nonumber \\
-(\Delta\varphi)_2=-\frac{2}{ab}\left(\frac{b\,\varphi_1+a\,\varphi_3}{a+b}
-\varphi_2\right)=\epsilon\,\varphi_2\,,
\nonumber \\
-(\Delta\varphi)_3=-\frac{2}{bc}\left(\frac{c\,\varphi_2+b\,\varphi_1}{b+c}
-\varphi_3\right)=\epsilon\,\varphi_3\,.
\label{ap1}
\end{gather}
For slowly varying from vertex to vertex variables $\varphi_i$
the system of equations (\ref{ap1}) transforms to the continual equation
$-\Delta\,\varphi=\epsilon\,\varphi$.
The three eigenvalues of Eq. (\ref{ap1}) are as follows:
\begin{gather}
\epsilon_1=0\,,
\nonumber \\
\epsilon_{2,3}=\frac{a+b+c}{abc}\left[1\pm\sqrt{1-\frac{8\,abc}{(a+b)(a+c)(b+c)}}\,\right]\,.
\label{ap2}
\end{gather}
If $a\rightarrow 0$ and $(a+b+c)=\const$, then
\begin{gather}
\epsilon_2\sim\frac{2(b+c)}{abc}\longrightarrow\infty\,, \qquad
\epsilon_3=\frac{4}{bc}\,.
\label{ap3}
\end{gather}
{\psfrag{Ph1}{\kern-5pt\lower-1pt\hbox{\large $\phi_1$}}
\psfrag{Ph2}{\kern0pt\lower0pt\hbox{\large $\phi_2$}}
\psfrag{Ph3}{\kern0pt\lower0pt\hbox{\large $\phi_3$}}
\psfrag{Ia}{\kern0pt\lower0.5pt\hbox{\large $a$}}
\psfrag{Ib}{\kern0pt\lower0.5pt\hbox{\large $b$}}
\psfrag{Ic}{\kern0pt\lower0.5pt\hbox{\large $c$}}
\begin{figure}[tbp]
 \includegraphics[width=0.20\textwidth]{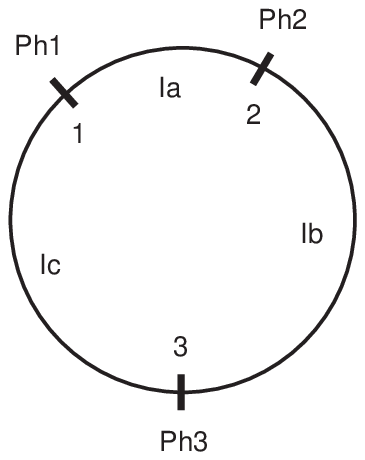}
\caption{}
\label{One}
 \end{figure}}


Consider the same problem for the discrete Laplace operator on a one dimensional
cycle with 4 vertexes separated in order by distances
$a$, $b$, $c$ and $d$. Then the eigenvalues of the operator satisfy
the following equation:
\begin{gather}
\epsilon^4-2\epsilon^3\left(\frac{1}{cd}+\frac{1}{bc}+\frac{1}{ab}+\frac{1}{ad}\right)+
\nonumber \\
+4\epsilon^2\left[\frac{1}{bc^2d}+\frac{1}{ab^2c}+\frac{1}{acd^2}+\frac{1}{a^2bd}+\frac{2}{abcd}-\right.
\nonumber \\
\left.-\frac{1}{c^2(b+c)(c+d)}-\frac{1}{b^2(a+b)(b+c)}-\right.
\nonumber \\
\left.-\frac{1}{a^2(a+b)(a+d)}
-\frac{1}{d^2(a+d)(c+d)}\right]-
\nonumber \\
-8\epsilon\left[\frac{1}{ab^2c^2d}+\frac{1}{abc^2d^2}+\frac{1}{a^2bcd^2}+\frac{1}{a^2b^2cd}-\right.
\nonumber \\
\left.-\frac{a+c}{ab^2cd(a+b)(b+c)}-\frac{b+d}{a^2bcd(a+d)(a+b)}-\right.
\nonumber \\
\left.-\frac{b+d}{abc^2d(b+c)(c+d)}-\frac{a+c}{abcd^2(a+d)(c+d)}\right]=0\,.
\label{ap4}
\end{gather}
Though the exact solution we do not obtained
the approximate solutions of this equation in two interesting here
special cases are written out:
\begin{gather}
b=d=l\,, \qquad a\rightarrow 0\,, \qquad c\rightarrow 0\,:
\nonumber \\
\epsilon_1=0\,, \quad \epsilon_2\approx\frac{4}{l^2}\,,
\nonumber \\
\epsilon_3\approx\frac{4}{la}\rightarrow\infty\,, \quad
 \epsilon_4\approx\frac{4}{lc}\rightarrow\infty\,.
\label{ap5}
\end{gather}
\begin{gather}
c=d=l\,, \qquad  a\rightarrow 0\,, \qquad b\rightarrow 0\,:
\nonumber \\
\epsilon_1=0\,, \quad \epsilon_{2,3}\approx\frac{2}{l(a+b)}\rightarrow\infty\,,
\nonumber \\
\epsilon_4\approx\frac{2}{ab}-\frac{4}{l(a+b)}\rightarrow\infty\,.
\label{ap6}
\end{gather}

\section{}

For clearness it is useful to see the phenomenon of imposing the
second class constraints without changing of quantum field
equations on an example of free Klein--Gordon theory. The
Klein--Gordon fields are expanded as follows:
\begin{gather}
\phi(x)=\sum_{\bk}\frac{1}{\sqrt{2\omega_{\bk}}}\left(
a_{\bk}\phi_{\bk}(x)+a_{\bk}^{\dag}\phi_{\bk}^*(x)\right)\,,
\nonumber \\
\pi(x)=-i\sum_{\bk}\sqrt{\frac{\omega_{\bk}}{2}}\left(
a_{\bk}\phi_{\bk}(x)-a_{\bk}^{\dag}\phi_{\bk}^*(x)\right)\,,
\nonumber \\
\omega_{\bk}=\sqrt{\bk^2+m^2}\,, \qquad
[a_{\bk},\,a_{\bp}^{\dag}]=\delta_{\bk\bp}\,.
\label{ap21}
\end{gather}
Here $\{\phi_{\bk}(x)\}$ is the complete set of orthonormal
functions, so that
\begin{gather}
\sum_{\bk}\phi_{\bk}(x)\,\phi_{\bk}^*(y)\bigg|_{x^0=y^0}=\delta^{(3)}(\bx-\by)\,,
\nonumber \\
\Delta\,\phi_{\bk}(x)=-\bk^2\,\phi_{\bk}(x)\,.
\label{ap22}
\end{gather}
The Hamiltonian
\begin{gather}
{\cal
H}=\int\d^3x\left(\frac12\pi^2+\frac12\nabla\phi\nabla\phi+\frac{m^2}{2}\phi^2\right)=
\nonumber \\
=\frac12\sum_{\bk}\omega_{\bk}\left(a_{\bk}a_{\bk}^{\dag}+a_{\bk}^{\dag}a_{\bk}\right)\,.
\label{ap23}
\end{gather}
Equations of motions are obtained with the help of Eqs.
(\ref{ap21})--(\ref{ap23}):
\begin{gather}
\dot{\phi}(x)=-i[\phi(x),\,{\cal H}]=
\nonumber \\
=-i\sum_{\bk}\sqrt{\frac{\omega_{\bk}}{2}}\left(
a_{\bk}\phi_{\bk}(x)-a_{\bk}^{\dag}\phi_{\bk}^*(x)\right)=\pi(x)\,,
\nonumber \\
\ddot{\phi}(x)=\dot{\pi}(x)=-i[\pi(x),\,{\cal H}]=
\nonumber \\
=-\sum_{\bk}\frac{\omega_{\bk}^2}{\sqrt{2\omega_{\bk}}}\left(
a_{\bk}\phi_{\bk}(x)+a_{\bk}^{\dag}\phi_{\bk}^*(x)\right)=(\Delta-m^2)\,\phi(x)\,.
\label{ap24}
\end{gather}

Now let us impose any number of pairs of second class constraints
\begin{gather}
a_{\bk_i}=0\,, \qquad a_{\bk_i}^{\dag}=0\,, \qquad
i=1,\,2\,\ldots\,.
\label{ap25}
\end{gather}
Then the sums $\sum_{\bk}$ in  (\ref{ap21}), (\ref{ap23}) and
(\ref{ap24}) transform to the reduced sums $\sum_{\bk\neq\bk_i}$.
Nevertheless equation of motion (\ref{ap24}) retains its canonical
form
\begin{gather}
\left(\partial^2/(\partial x^0)^2-\Delta+m^2\right)\,\phi(x)=0\,.
\label{ap26}
\end{gather}
The dynamical reason for this conclusion in considered example is
that the commutators of the constraints (\ref{ap25}) with
Hamiltonian are proportional to the constraints, i.e. they are
equal to zero in a weak sense:
\begin{gather}
[a_{\bk_i},\,{\cal H}]=\omega_{\bk_i}a_{\bk_i}\,, \qquad
[a_{\bk_i}^{\dag},\,{\cal H}]=-\omega_{\bk_i}a_{\bk_i}^{\dag}\,.
\label{ap27}
\end{gather}
Therefore, as it was shown in Section 3, equations of motion in
reduced theory must retain their canonical form.

\end{document}